\documentclass[aps,draft,12pt,floatfix]{revtex4-1}
\usepackage{mathtools}
\usepackage{latexsym}
\usepackage{graphicx}
\usepackage{dcolumn}
\usepackage{bm}
\usepackage{amssymb}
\usepackage{amsmath}
\usepackage[space]{grffile}

\usepackage{subfig}

\begin{document}

\newcommand{\niceref}[1] {Eq.~(\ref{#1})}
\newcommand{\fullref}[1] {Equation~(\ref{#1})}
\title{Quantum quench dynamics}

\date{\today}

\begin{abstract}
Quench dynamics is an active area of study encompassing condensed matter physics and quantum information, with applications to 
cold-atomic gases and pump-probe spectroscopy of materials. Recent theoretical progress in studying quantum quenches is reviewed. 
Quenches in interacting one dimensional systems as well as systems in higher spatial dimensions are covered. The appearance of 
non-trivial steady states following a quench in exactly solvable models is discussed, and the stability of these states to 
perturbations is described. Proper conserving approximations needed to capture the onset of thermalization at long times 
are outlined. The appearance of universal scaling for quenches near critical points, and the role of the 
renormalization group in capturing the transient regime, are reviewed. Finally the effect of quenches near critical points on the 
dynamics of entanglement entropy and entanglement statistics is discussed. The extraction of critical exponents from the entanglement 
statistics is outlined. 
\end{abstract}

\author{Aditi Mitra}

\affiliation{Department of Physics, New York University, 4 Washington Place, New York, NY, 10003, USA}
\maketitle

\tableofcontents
\newpage
\section{Introduction} \label{intro}

The quantum quench studies how a many particle system
prepared initially in the ground state of a Hamiltonian $H_i$, evolves unitarily in time
following the sudden change of the parameters to a final Hamiltonian
$H_f$\cite{Calabrese05b,Calabrese2006}. Dynamics generated by a quantum quench has become an active topic of research because it poses many
fundamental questions that can also be studied by current generation experiments.

Among experimental systems, the most striking are cold-atomic gases trapped in tunable optical lattices
which can realize ideal Hamiltonians~\cite{Bloch08,Cazalillarev11}. In addition, the
parameters of the Hamiltonian, such as interaction strength between
particles, and lattice parameters, can be tuned rapidly in time.
Being well isolated from the surroundings, it is a good approximation to assume the absence of an external thermalizing bath.
In fact these many particle systems
are usually far from equilibrium, with long thermalization times. This is particularly true for
cold atoms in one dimensional ($1d$) optical lattices where thermalization
times can be unobservably long~\cite{Weiss06,Gring12,Trotzky12,Bloch15}.

Some questions that naturally arise in the field of quantum quenches, and that have the scope of being tested experimentally are,
what are the mechanisms and time-scales for thermalization of closed quantum systems~\cite{Deutsch91,Srednicki94,Rigol08}?
What is the difference in the dynamics of generic interacting systems and special Hamiltonians known to be 
integrable~\cite{Andrei12,Iyer13,Mussardo13,Wouter16}?
Since degrees of freedom are not counted the same way in classical and quantum systems, how does
the notion of integrability generalize to quantum systems~\cite{Yuzbashyan11}? Are there any quantum generalizations to the
Kolmogorov-Arnold-Moser (KAM) theorem~\cite{KAM,Konik15}? How does the phenomena of many body localization~\cite{Anderson58,Basko06,Oganesyan07},
often described by the emergence of quasi-local integrals of motion~\cite{Huse15}, depend on spatial dimensions, range of interactions
and disorder?

Besides the formal notion of integrability,
from the condensed matter perspective, natural questions that also arise is if the initial Hamiltonian or final Hamiltonian or both,
can support collective order
such as ferromagnetism or superconductivity or for that matter topological order, how does this order respond to or develop under a quench? Is
there some universality in the dynamics when the quench is in the vicinity of a critical point? If so can one generalize powerful
theoretical methods such as the renormalization group (RG) to study quantum quenches?

In attempting to understand how thermalization occurs in closed quantum systems, a useful quantity to study is the reduced density matrix.
This is obtained from the full density matrix by tracing or integrating out some subset of the full Hilbert space.
While the full density matrix can never show thermalization under unitary time evolution, the reduced density matrix can look
effectively thermal with the remaining (integrated out) system acting like an effective reservoir for it.
Understanding the dynamics of the reduced density matrix in the context of quantum quenches has in turn lead to a
synergy between two different fields, that of condensed matter and quantum 
information~\cite{Vidal03,Preskill06,Levin06,Haldane08,Casini09,Eisert10}.
Questions that interest both communities are how the entanglement entropy evolves in time~\cite{Calabrese2007,Calabrese07b,Hartman2013}?
What features of the entanglement entropy and entanglement
statistics can differentiate between ergodic and non-ergodic systems~\cite{Lin12,Huse13}? The study of how entanglement scales with size
of the sub-system,
and with time after the quench, has in turn lead to the development of novel numerical and variational methods for studying quench
dynamics~\cite{TDMRGrev11,Cirac11,Altman17}.

The discussion above shows that the field of quantum quenches is very broad, touching upon many concepts.
It is difficult to do justice to all these topics in their entirety.
This review will discuss some of these topics, with emphasis based on the author's own research contribution in the field.

The outline of the paper is as follows. To orient the readers, in Section~\ref{model},
quantum quenches in some exactly solvable models of free fermions and free bosons
in $1d$ will be presented. The key features of the resulting nonequilibrium steady states will be discussed. The
advantage of studying quenches in $1d$ is that even effectively free theories capture non-trivial correlations~\cite{Giamarchibook}.
This can in turn help build intuition about dynamics of correlated states in higher spatial dimensions.

In Section~\ref{1d} various perturbations around the free limit will be considered. The perturbations will be non-linear
in a way as to break the underlying integrability.
The emphasis will be on developing methods that are general to non-integrable systems and also applicable to higher spatial dimensions.
In Section~\ref{1da} it will be shown  that even in the presence of integrability breaking non-linear terms, it is helpful to think
of the dynamics in terms of an intermediate time prethermal (or collisionless) regime~\cite{Berges04,Kehrein08,Kollar11,Marcuzzi13}.
It will be shown that for quenches to a critical point, universality emerges in the prethermal dynamics, and
an RG approach can be used to study its properties.

Finally the long time collisional or inelastic scattering dominated regime will be discussed in Section~\ref{1db}. Here it will
be outlined how a two-particle irreducible formalism can be used to obtain quantum kinetic equations that preserve conservation laws.
Thermalization times obtained from the quantum kinetic equation for a $1d$ system with interactions and weak disorder will be discussed.

Following this, in Section~\ref{hd} results will be presented in higher spatial dimensions $d>2$ where
a quench to a critical point will be considered.
Universality in the dynamics will be highlighted and the success of RG in capturing
key features of the scaling will be discussed.
In Section~\ref{hde}, the time evolution
of the entanglement entropy and entanglement statistics for the critical quench in $d>2$ will be discussed, and
how universal physics may be extracted from
the entanglement statistics will be described.
Finally in Section~\ref{conclu} we present our outlook.

\section{Quenches in $1d$ effectively free theories: nonequilibrium steady states}\label{model}

In this section we will consider quenches in free theories to highlight the new phenomena that can arise due to a quench.
We will consider a spatially inhomogeneous quench, a homogeneous quench, and finally a quench in a system with interactions
and disorder.

\subsection{Inhomogeneous quench}

Let us first study a spatially inhomogeneous quench in a simple model of non-interacting fermions in $1d$.  
We will consider the spin 1/2 Heisenberg $XX$ chain, which, after a Jordan Wigner transformation, maps to free spinless fermions
with nearest neighbor hopping. We will show that even this simple model can exhibit some remarkably interesting nonequilibrium states
generated by a quench.

Imagine initially applying a magnetic field along the $\hat{z}$ direction, that
varies linearly along the chain, changing sign at the center. Thus the initial Hamiltonian is
\begin{eqnarray}
H_i = -J\sum_{j}\biggl[S^x_j S^x_{j+1}+ S^y_j S^y_{j+1}-\frac{j Fa}{J}S^z_j\biggr],
\end{eqnarray}
where $j$ is the position index, $a$ is the lattice spacing, and the magnetic field varies linearly as $j F a$.
The ground state of $H_i$, which will act as the initial state for the quench, is a domain wall centered at $j=0$, and connects regions
with maximum positive and negative polarization along $+\hat{z}$ on opposite ends of the chain.
The quench will be the sudden switch off of the magnetic field so that
the time evolution is simply due to the homogeneous $XX$ chain.

After the quench, the domain wall is no longer a stable state, and acquires a time evolution.
In the language of fermions, the initial state has a density imbalance due to a spatially varying chemical potential. The quench involves the switching
off of the chemical potential so that the density imbalance is no longer stable, and as time evolves, there is flow
of particles from one side of the chain to the other. In the language of spins, there is a flow of spin current from one side
of the chain to the other. This current causes the domain wall to broaden ballistically with a velocity set by the lattice
parameters~\cite{Antal99,Platini07,Lancaster10}.

We take the length $L$ of the chain to be the largest scale in the problem. Effectively this involves studying the dynamics before the width of
the domain wall becomes comparable to the system size.
We find that the transverse
spin correlation function at a time $t$ after the quench, defined as,
\begin{eqnarray}
C_{xx}(j,j+n,t) =\langle \Psi_i|e^{i H_ft}S^x_{j}S^{x}_{j+n}e^{-i H_f t}|\Psi_i\rangle,
\end{eqnarray}
reaches a nonequilibrium steady state of the form~\cite{Lancaster10}
\begin{eqnarray}
C_{xx}\biggl(j,j+n, \frac{2vt}{na}> 1\biggr) \rightarrow C_{xx}^{\rm eq}(n)\cos\left(\frac{2\pi n}{\lambda}\right).\label{dwtw}
\end{eqnarray}
$j,j+n$ correspond to points in the central region of the chain where the magnetization has dropped to zero.
$C_{xx}^{\rm eq}$ is the ground state correlation of the uniform $XX$ chain, and falls off
as a power-law in position, $C_{xx}^{\rm eq}\sim \frac{1}{\sqrt{n}}$~\cite{Ovchinnikov07}. The only knowledge the system has of the
initial state is through the spatially oscillating prefactor of wavelength $\lambda$.  It's physical
origin is the spin current flowing through the region, and corresponds to spins twisting in a spiral pattern in the $XY$ plane,
$S^+_j = S^x_j+i S^y_j\sim S_0^+e^{2\pi i j/\lambda}$.
The strength of the spin current is proportional to the twist rate of the in-plane spins, and
depends on the polarization $\pm m_0$ at the two extremes
of the domain wall  as, $\lambda = \frac{2}{m_0}a$.
In the example discussed above, the domain wall polarization at the extremes are maximal $m_0=1/2$, so that
the wavelength of the oscillation is $\lambda= 4 a$.

Not just domain walls, but other kinds of initial density inhomogeneities relax via the creation
of such transient current carrying states. For hard core bosons released from a trap, for example, they imply appearance of
quasi-condensates at finite momentum~\cite{Rigol04,Vidmar15}.

There are two interesting questions one can address here. One is that in $1d$, powerful
methods such as bosonization work well in capturing the low energy properties. How well do these
methods capture the quench dynamics? Secondly, how do non-trivial interactions such
as $J_z \sum_iS^z_iS^z_{i+1}$ affect the results above~\cite{Langer09,Foster11,Santos11,Lancaster10b,Vidmar16,Bertini16,Doyon16}?
In particular, how does the domain wall
expansion velocity change? How is the steady state spin current affected? The last question is related to
transport in interacting models, where bounds on the optical conductivity,
such as Mazur's inequalities exist~\cite{Mazur69}. How are these bounds, established for systems in thermal
equilibrium, affected by a quench?

Here we will only address what features of the nonequilibium state above is captured by bosonization.
Within a bosonization picture,
one may encode the effect of forward scattering interactions from non-zero $J_z$ into
the Luttinger parameter $K$, while the back-scattering interaction are captured by a
non-linear interaction in the form of a cosine field. In the phase where back-scattering is RG irrelevant, the homogeneous
$XXZ$ chain is~\cite{Giamarchibook}
\begin{eqnarray}
H_{\rm LL} = \frac{u}{2\pi}\int dx \biggl[K(\partial_x\theta)^2+\frac{1}{K}(\partial_x \phi)^2\biggr].\label{eq1}
\end{eqnarray}
The domain wall density is encoded in the bosonic variable $\partial_x \phi$,
while the current density is in the variable, $\partial_x \theta$, where $\phi, \frac{1}{\pi}\partial_x \theta$ are canonically conjugate.
We adopt a convention where Luttinger parameter $K=1$ is $J_z=0$ or non-interacting fermions,
while $K<1$ are repulsive fermions and $K>1$ attractive fermions.

In order to compare results with the spin chain we modify $H_{\rm LL}$ above by introducing a spatially
varying chemical potential or magnetic field,
and the quench involves switching it off.
Studying domain wall dynamics using bosonization reveals
that indeed a steady state emerges as the domain wall broadens.
This state has the same property as in Eq.~\eqref{dwtw}, namely the correlator has
the same power-law correlations $1/r^{1/(2K)}$ as in the ground state of $H_{\rm LL}$, but modulated in space due to the
resulting current carrying state $\partial_x\theta \propto 1/\lambda$. The associated
spiraling in-plane component of the spin has a wavelength~\cite{Lancaster10},
\begin{eqnarray}
\lambda = \frac{2K}{m_0}.
\end{eqnarray}

Thus bosonization does manage to capture the main features, although, as expected it
does not capture the detailed microscopics such as precise domain wall velocity
(in the continuum there is only one velocity, whereas on the lattice
there are a range of velocities), and also the detailed profile of the domain wall~\cite{Eisler13,Evertz15}.
The above dependence of $\lambda$ on the interaction parameter $K$ or equivalently the anisotropy $J_z$ obtained
from bosonization, was also recovered in numerical studies provided the
currents (or equivalently the domain wall polarization) was not too large~\cite{Misguich13}.

The situation described above cannot be probed at very long times numerically as eventually the domain wall width will reach the system
size. Thus interesting questions such as how the magnitude of the current depends on irrelevant operators and
inelastic scattering remain unanswered. Carrying out such a study analytically is also an open problem.

\subsection{Homogeneous quench}\label{1dhom}
A well studied nonequilibrium steady state is that arising after an interaction quench from $K_0 \rightarrow K$ in the Luttinger
liquid~\cite{Cazalilla06,Iucci09,Mitra11} and lattice models~\cite{Karrasch12,Kennes13,Collura15}.
Thus the initial state is the zero temperature
ground state of free bosons with interaction parameter $K_0$, while the time evolution is with respect to Eq.~\eqref{eq1}.
To preserve Galilean invariance, we will also impose that the initial and final velocities change as follows $u_0K_0\rightarrow u K$.

The initial and final Hamiltonians are diagonalized by two different sets of bosonic operators that are related to each
other by a Bogoliubov transformation. This transformation, together with the initial state, completely determine
the occupation probabilities of the bosonic modes of the final Hamiltonian.  Clearly there is no thermalization in a Gaussian theory,
but only relaxation to a steady state, the latter being well described by a
Generalized Gibbs Ensemble (GGE)~\cite{Rigol07} that accounts for
the conservation of population of the bosonic modes. For the example being discussed, the steady state
continues to be a power-law but with a new exponent
that depends on the initial and final Luttinger parameters.
In particular,
\begin{eqnarray}
&&C_{\phi\phi}(x,t) =\langle\psi_i |e^{i H_f t}\cos(\phi(x))\cos(\phi(0))e^{-iH_ft}|\psi_i\rangle, \label{Cdef}\\
&&\underrightarrow{t\rightarrow \infty, L\rightarrow \infty}\,\,\,\, \frac{1}{x^{2K_{\rm neq}}},
\end{eqnarray}
where $K_{\rm neq}= \frac{1}{8}K_0\left(1+\frac{K^2}{K_0^2}\right)$.
This power-law decay in position is faster than in the ground state of the final Hamiltonian, the latter following
from $K_{\rm neq}$ by setting $K_0=K$. Yet the post quench steady state is not a thermal state as for the latter the decay would be exponential
in position. A popular way to describe this is via a mode-dependent temperature. Note that, despite the power-law, the steady state is not
described by a new Luttinger
parameter as the duality between the $\phi$ and $\theta$ correlators, a hall mark of the Luttinger liquid, is lost in this
nonequilibrium steady state~\cite{Schiro14,Schiro15}.

We should also mention that quenches for free bosons with a gapless spectrum
do not always generate power-laws, even when the initial state is the zero temperature ground state.
We highlight this in the next sub-section via an example where we switch on a disorder potential as a quench.

An active area of study which we will not cover is to generalize the GGE to integrable models that cannot be written
as free bosons or free fermions. Much progress has been made in this direction where 
GGEs have been constructed that capture the conservation of non-local operators~\cite{Caux13,Ilevski15,Caux16}.

\subsection{Disorder quench}
An interesting question concerns the modification of the above results for free and clean systems, by disorder. To address this, 
we start with an initial Hamiltonian corresponding to Eq.~\eqref{eq1}, and switch on disorder. In the language of bosonization,
the disorder potential can be split into a forward scattering part and a backward scattering part~\cite{Giamarchibook}.
The latter makes the problem non-linear,
and its effect will be discussed later. For now let us consider only the effect of random forward scattering. Thus the final Hamiltonian
is
\begin{eqnarray}
H_f = H_{\rm LL} -\frac{1}{\pi} \int dx \eta(x) \partial_x \phi(x),
\end{eqnarray}
where $\eta$ is a Gaussian distributed disorder $\overline{\eta(x)\eta^*(x')}=D_f \delta(x-x')$.
In equilibrium, one may completely remove the effect of forward scattering by a simple change of variables
$\phi(x) \rightarrow \phi(x) -\frac{K}{u}\int^x dy \eta (y) $. Thus forward scattering does not affect the two point correlations.
In contrast for a quench, the initial state being the ground state of the fields before the shift, and thus a highly excited state for
the shifted fields, the averages with respect to the initial state now do depend on the forward scattering disorder potential.

In a semiclassical picture, the quench creates excited quasiparticles that when propagating through the system pick
up random phases associated with the disorder potential. Before averaging over disorder, the correlators decay with the
same power-law as in the ground state of the clean system $H_{\rm LL}$, but in addition carry random phases. For example,
the $C_{\phi\phi}$ correlator has the following form before disorder averaging~\cite{Tavora14},
\begin{eqnarray}
&&C_{\phi\phi}(x,t) = C_{\phi\phi}^{\rm eq}(x) e^{-i\frac{K}{u}\sum_{\epsilon=\pm}\biggl[\int_x^{x+\epsilon u t} dy\eta(y)-\int_0^{\epsilon u t} dy\eta(y)\biggr]}.
\end{eqnarray}
$C_{\phi\phi}^{\rm eq}$ is the ground state correlator of $H_{\rm LL}$. The accompanying phase factors
show that the operator at position $x$ and a time $t$ after the quench acquires contributions from random phases accumulated
from the region $[x, x+u t]$ for left moving quasiparticles, and the region $[x,x-ut]$ for right moving quasiparticles.

On disorder-averaging this result,
the random phases lead to an exponential decay of the correlations. There is an interesting light-cone behavior where for
$nut< r$ the disorder averaged correlators decay exponentially in time, while for $n ut > r$, a crossover to a steady state
with exponential decay in position occurs. Note $n=1$ or $2$ depending on the correlator being studied. In particular $n=1$
for $C_{\phi\phi}$, while $n=2$ for the correlator for the conjugate operators, $C_{\theta\theta}$~\cite{Tavora14}.

Thus, even though there is no inelastic scattering and hence no generation of a temperature, the correlators decay exponentially
in position at long times. This is an example of elastic dephasing. In subsequent sections we will discuss the effect of
non-linearity arising from the back-scattering potential. We will discuss its effect on a quench when it is
irrelevant and marginally irrelevant.

\section{Quenches in $1d$ interacting field theories}\label{1d}
In this section we discuss the effect of non-linear perturbations on the free fermion and free boson steady states discussed above.
We will consider the effect of the non-linearities for the cases where they are RG irrelevant or marginal, as this will allow
us to carry out controlled perturbative schemes. Various non-linearities are of interest. We will primarily consider the
role of a back-scattering interaction. These could arise due to a local impurity potential, an underlying lattice, and
even a disorder potential.

\subsection{Universal scaling in prethermal regime}\label{1da}

Consider an initial state which is the ground state of the Luttinger liquid with an interaction parameter $K$ as described by Eq.~\eqref{eq1}.
The quench corresponds to the sudden switch on of a non-linear term corresponding to back-scattering from a commensurate lattice. Thus the final
Hamiltonian is,
\begin{eqnarray}
H_f=H_{\rm LL}+V;\,\, V= -g\int dx \cos(2\phi).
\end{eqnarray}
The prequench Hamiltonian is described by free bosons, while the post quench Hamiltonian is the celebrated sine-Gordon model.
Note that many different lattice models can lead to the same continuum field theory. For example, the post quench Hamiltonian
could represent both a spin-chain, as well as the Bose Hubbard model. This may lead to confusion as one model is integrable, and the
other is not. In the phase where the cosine term is irrelevant, the differences in the two models show up at the level of how the
lattice or UV physics is introduced.

The model above has a quantum critical point. For the spin chain the critical point separates the gapless $XX$ phase
from the gapped Ising phase, while for the Bose Hubbard model, it separates the superfluid phase from the Mott insulating phase.
The transition between the two phases is of the Berezenskii-Kosterlitz-Thouless (BKT) kind. For $g\rightarrow 0$, the
critical point is located at $K=2$, with $K<2$ being the gapped phase where the cosine potential is RG relevant. In contrast for $K>2$,
the cosine potential is RG irrelevant and the spectrum stays gapless~\cite{Giamarchibook}.
At the critical point, the system in the zero temperature ground state shows UV independent universal physics. A cautionary note here
is that the universality is not in all quantities, but some special set of correlators directly related to 
the ``order-parameter''~\cite{Singh89,Affleck98}.

We study quenches in the gapless phase where the cosine term is
irrelevant and at the critical point where it becomes marginally irrelevant, perturbatively in the cosine potential.
We show that when the parameters of $H_f$ are tuned to be at the critical point, an intermediate time
prethermal phase exists which shows UV independent universal physics.

The two point correlation functions which show the onset of universal behavior in equilibrium are defined in Eq.~\eqref{Cdef}. For the spin
chain it represents the staggered anti-ferromagnetic $(-1)^nS^z_iS^z_{i+n}$ correlation function, while for the Bose Hubbard model
it measures the superconducting phase fluctuations.
We study how $C_{\phi\phi}$ evolves in time perturbatively in the cosine potential and find that
when $K=2$, the leading non-zero correction grows logarithmically as follows (below
we set velocity $u=1$, position and time are in units of a lattice scale, and we give results only
for macroscopic position and times $r,t\gg 1$)~\cite{Mitra13a},
\begin{eqnarray}
&&C_{\phi\phi}(r,t) = C_{\phi\phi}^0(r)\biggl[1+\pi g \biggl(\ln r^2 +  \ln t^2 - \ln\sqrt{64 t^2+(4t^2-r^2)^2}\biggr)\biggr].
\end{eqnarray}
In the absence of the cosine potential ($g=0$), this correlator decays as $C_{\phi\phi}^0(r) \sim \frac{1}{r}$.

The above shows light-cone dynamics where, when $r >> 2t $, the correlator decays in position primarily as in the initial state.
Since our initial state was gapless, the correlator at these short times has a power-law in position decay. Under the effect of the
cosine, this result is corrected by
logarithms that grow in time as $C_{\phi\phi}(r\gg 2t)\sim \frac{1}{r}\biggl[1+2\pi g \ln t\biggr]$. In the
opposite limit of long times $r \ll 2t$, the logarithmic correction is cut off by the distance so that
$C_{\phi\phi}(r\ll 2t)\sim \frac{1}{r}\biggl[1+2\pi g \ln r\biggr]$. However right on the light-cone,
$r=2t$, the logarithmic corrections are qualitatively different. Quite generically one expects that
on the light cone the correlators will be enhanced due to contributions from
ballistically propagating, and initially entangled quasiparticles reaching the the positions $0,r$ simultaneously.
Our result indicates that for a quench to the critical point this enhancement appears as a further enhancement of
the logarithmic correction $C_{\phi\phi}(r= 2t)\sim \frac{1}{t}\biggl[1+2\pi g \left(\frac{3}{2}\right)\ln t\biggr]$,
note the change in the prefactor of the logarithm from $2\pi g \rightarrow 2\pi g (3/2)$.

Being perturbative, we do not expect these results to be valid at long times. Nevertheless,
the UV independence of the perturbative treatment indicates that there is an intermediate time regime where
an RG analysis may be performed to capture the quench dynamics.

As in equilibrium, we employ a Wilson Fisher approach to gradually integrate out short wavelength degrees
of freedom, and  study how parameters renormalize. For the case under discussion where
a periodic potential was quenched, the couplings $g,K$ obey the same RG equations as in
equilibrium, which are of the BKT type.
The difference now is that the RG flows have to be cut-off by ${\rm min}(2t,r)$, as is clear from
the perturbative results for the correlation function presented above. In addition, the correlations themselves
can be shown to obey a Callan-Szymanzik equation that relates correlators at long distances and/or times after the quench with
the short distance and short time correlators of a weak coupling problem, where for the latter a perturbative treatment can be used.

Employing such a procedure, we arrive at the following results~\cite{Mitra13a},
\begin{equation}
C_{\phi\phi}(r,t) \sim \begin{cases*}\frac{\sqrt{\ln t}}{r} & if $2t < r$\\
\frac{\ln t}{t} & if $2t = r$\\
\frac{\sqrt{\ln r}}{r} & if  $2t > r$
\end{cases*}
\end{equation}
The last long time result is the same as in the ground state of $H_f$~\cite{Singh89,Affleck98}. At this level of perturbation theory, our calculation
does not account for the long time limit where the energy injected by the quench will lead to inelastic scattering.
The above results show that a prethermal regime exists where although full scale inelastic scattering is absent, yet dephasing
effects from interactions
can strongly modify the correlators from a naive free fermion or free boson model.

\subsection{Thermalization via multi-particle scattering}\label{1db}
Eventually for a generic nonintegrable system thermalization sets in. When carrying out the RG, the indication
of the long time inelastic regime appears as
the generation of additional terms in the Keldysh action~\cite{Mitra11,Mitra12a,Mitra12b}.
For bosons, if $\phi_q$ is the quantum field, and $\phi_c$, the classical fields~\cite{Kamenevbook2011,BergesRev},
the new terms generated by the RG are of
the kind $\phi_q\partial_t \phi_{c}$ and $\phi_q^2$. The former corresponds to dissipation, while the latter to
an effective temperature.

Thus although the full system evolving unitarily will never show thermalization, the RG shows that the integrated out short wavelength
modes can act as an effective reservoir for the long wavelength modes. When the system is out of equilibrium, for example due to a quench,
this reservoir is where the long wavelength modes dissipate their energy. It is important to note that
dissipative terms are generated even when the interactions
are RG irrelevant indicating that irrelevant operators are responsible for thermalization to set in,
and therefore cannot be neglected at long times.

In order to study the full dynamics of this thermal phase, we develop a quantum kinetic equation approach.
N\"{a}ive kinetic equations do not work in $1d$ as two particle scattering cannot relax a distribution function.
In addition for our Hamiltonian,
the non-linear term is not $\phi^4$ but is more complex, being proportional to $\cos\phi$. Moreover, when $\cos\phi$
is RG irrelevant, the $\phi$ field has strong fluctuations, and therefore one may not Taylor expand $\cos\phi$.

We adopt a two particle irreducible (2PI) formalism
that allows one to treat generic interactions such as the cosine term, and also allows us to make
conserving approximations. The principle is the following.
One writes a Keldysh field theory or action $e^{i\Gamma}$ as a functional $\Gamma(G)$ of two particle averages,
the Green's function  $G$.
The solution for $G$ is
the saddle point of this field theory $\delta \Gamma/\delta G=0$. It can be shown that the saddle point condition is equivalent to the
Dyson equation in Keldysh space~\cite{CornwallJackiwTomboulis,Nishiyama,AST61,BergesRev}.  
Next, one makes suitable approximations for the 2PI action, that translates to
approximations for the self-energy, Dyson equations, and consequently the Green's functions.

Conservation laws follow naturally in the 2PI approach
from Noether's theorem. For example deviations that correspond to globally conserved quantities do not change the action.
On the other hand, local
variations of these conserved quantities change the 2PI action. By requiring that these local changes occur around a saddle point,
and therefore vanish to leading order, one may derive hydrodynamic equations. Thus one may obtain consistent definitions for the
stress momentum tensor if energy and momentum are conserved, and the diffusion equation, if total number is conserved.
Any approximations made to the 2PI action translate to consistent approximations to the Green's functions,
the hydrodynamic equations, and conserved densities~\cite{Ivanov98}.

We carried out this procedure for the Luttinger liquid with a cosine potential. We studied the case
when the cosine was due to an underlying regular lattice~\cite{Tavora13} and also when it was due to
a disorder potential~\cite{Tavora14}. We briefly discuss the results for the case of disorder. The effect of the forward scattering
disorder has already been discussed in the previous section. The backward scattering disorder can be written as,
\begin{eqnarray}
V_{\rm dis} = \int dx \biggl[\xi^*e^{2i\phi} + \xi e^{-2i\phi}\biggr],
\end{eqnarray}
where we assume the disorder to be Gaussian distributed $\overline{\xi(x)\xi^*(x')}={\cal D}_b\delta(x-x')$.
The ground state of $H=H_{\rm LL}+V_{\rm dis}$ is well studied, showing a BKT transition from the
superfluid to the Bose-glass phase at Luttinger parameter close to $K=3/2$~\cite{Giamarchibook}. We study quench dynamics
when the disorder is suddenly switched on. The forward scattering disorder causes dephasing, while
backward scattering causes inelastic scattering.

Under conditions that this cosine potential is RG irrelevant,
we constructed the 2PI action to second order in the disorder potential.
Due to the interaction vertex being of the form $e^{2i\phi}$, a key
feature of the kinetic equation is that, even the second order correction allows for multi-particle
scattering between bosons as the exponential keeps track of all powers of the $\phi$ field.

In solving the kinetic equation, we can make further approximations such as keeping only the leading
order Wigner expansion, making the kinetic equation local in time. Some very
general results can then be obtained for thermalization times, and how it is affected as one approaches the superfluid-Bose
glass quantum critical point. We discuss some key results below.

In general the relaxation of the initial out of equilibrium population is not a pure exponential.
Nevertheless by matching the long time behavior of the kinetic equation with an exponential relaxation rate,
we found the thermalization time $t_{\rm th}$ to be $t_{\rm th}^{-1}\sim {\cal D}_b \biggl[T_{\rm eff}\biggr]^{K-1}$, where
${\cal D}_b$ is the strength of the backscattering potential, while
$T_{\rm eff}$ is a measure of density of excited quasiparticles generated by the quench. In particular, the bosonic density
at long wavelengths $\hbar/p$, is $n_p=T_{\rm eff}/(u|p|)$ where $T_{\rm eff} = {\cal O}\left({\cal D}_b,{\cal D}_f\right)$.
The dependence on ${\cal D}_{f,b}$ reflects
the fact that the sudden switch on of the disorder creates a non-zero density of bosonic excitations.

The above result for $t_{\rm th}$ shows that on
decreasing $K$ from a large value $K\gg 1$ towards the critical point
located at $K=3/2$, the relaxation rate increases.
This indicates that thermalization
"critically speeds up" as a quantum critical point is approached from the superfluid side. This is a very general result that occurs when
the system is out of equilibrium, and when the leading non-linearity responsible for thermalization is also
the one driving the phase transition. This result for ``critical speeding up'' is also
borne out in experiments~\cite{Grams14}.

The thermalization dynamics captured by a kinetic equation, as described above, effectively captures local equilibration.
Reaching global equilibrium is however more difficult as conserved quantities have to be transported over long distances.
Setting up a hydrodynamic approach to study this long time limit is an important open question. The hydrodynamic regime is
particularly intriguing in $1d$ as simple RG arguments~\cite{Lux13} indicate that non-linearities
of the Kardar-Parisi-Zhang (KPZ) type~\cite{KPZ86} are relevant and will modify the power-law associated with the hydrodynamic long time tails
from their na\"{i}ve Gaussian predictions~\cite{Henk12}.
Studies performed on the $1d$ Bose-Hubbard model indicate that such a regime may be identified numerically~\cite{Lux13}
where the long time averages of various observable were found to fall off in time
as $t^{-1/2}$ (the Gaussian prediction) with subleading corrections  proportional to $t^{-3/4}$ (whose origin is consistent
with KPZ scaling). Further analytic and numerical studies exploring this physics is an important direction of research.

\section{Quenches in spatial dimension $d>2$}\label{hd}

The methods developed above for $1d$ systems are rather general, and can be used to study quenches in higher spatial
dimensions. In this section we discuss quenches in a bosonic $\phi^4$ theory in general dimension $d$.
Time dependent mean-field treatments of various models have indeed shown the existence of dynamical phase transitions where
by tuning the parameters of the quench, the system at long times can go from exhibiting disordered to ordered behavior, with the transition
between the two characterized by the vanishing of an effective mass~\cite{Schiro10,Gambassi2011,Sciolla2013,Sondhi2013,Smacchia2014}.
Our goal is to uncover universal physics in the quench dynamics when the quench is tuned right at
the critical point~\cite{Sondhi2013,Chiocchetta2015,Maraga2015,MitGam16,Gambassi17}. In $1d$
there is no finite temperature critical point for a continuously broken symmetry. This limits the possibilities of
seeing universal physics in quenches because a quench injects an effective temperature into
the system, which cuts-off all power-law correlations. One may see some power-laws persist at intermediate times
as discussed in Section~\ref{1da}, but in $1d$ since the critical points are often of the BKT kind, interactions
only give weak logarithmic corrections.
We should emphasize that by universal we mean physics that is independent of microscopic
details of the system, such as interaction strength. By this definition it excludes power-laws such as those discussed
in Section~\ref{1dhom}, as these power-law decays depend on
the microscopic details via the Luttinger parameter.

The quench we study is one where initially the system is in the ground state of a bosonic system with a large mass. Thus
the initial state has short range correlations that decay within a few lattice sites.
The quench corresponds to reducing the mass from this large value to a very small value such that the final Hamiltonian is
tuned close to criticality. The interactions are always present, but if the initial mass is sufficiently large,
the effect of interactions on the initial state
is negligible. Thus we may model the initial state as the ground state of a free boson Hamiltonian $H_i$ with mass $\Omega_0^2$,
while the time evolution is due to an interacting Hamiltonian $H_f$ with mass $r \ll \Omega_0^2$ and interaction strength $u$,
\begin{eqnarray}
&&H_i = \int_{\vec{x}}\left[ \frac{1}{2}\vec{\pi}^2 + \frac{1}{2}(\nabla \vec{\phi})^2 + \frac{\Omega^2_0}{2}\vec{\phi}^2\right],\nonumber\\
&&H_f = H_i(\Omega_0^2 \rightarrow r) + \frac{u}{4!N}\int_{\vec{x}}\left(\vec{\phi}^2\right)^2,
\end{eqnarray}
above the fields are vectors $\vec{\phi}(\vec{x})=(\phi_1,\ldots,\phi_N)$ with $N$
components and  the model has an $O(N)$ symmetry. For $N=2$ this model captures the transition from a Mott insulator
to a superfluid phase, while for $N=3$ it describes a transition from a paramagnet to an antiferromagnet.

A quench injects energy into the system which translates into an effective temperature. This is seen even at the level
of free bosons ($u=0$), where the density of bosonic excitations
at momentum $k$ are $n_k = \frac{T_{\rm eff}}{\omega_k}$ with $\omega_k =\sqrt{k^2 + r}$ and $T_{\rm eff}=\Omega_0/4$. This temperature causes
the system to approach a finite temperature classical (rather than the zero temperature quantum) critical point.
Yet the time evolution towards such a state
is itself unitary, so that the full quantum dynamics needs to be accounted for.

On studying the time evolution perturbatively in the interactions, logarithmic corrections are found
at the classical upper critical dimension $d=4$.
An RG allows us to resum these logarithms and obtain scaling forms valid for general $\epsilon = 4 -d$, provided $d>2$ ($d=2$ being the lower
critical dimension of the theory)~\cite{Chiocchetta2015,MitGam16}.
The RG results can be benchmarked against exact solutions obtained from solving the Hartree-Fock equations in the $N=\infty$
limit~\cite{Maraga2015}.

A critical quench is characterized by an effective mass $r_{\rm eff}(t) = r + \frac{u}{6}\langle \left[\vec{\phi}(t)\right]^2\rangle$
which approaches zero at long times. For the model under discussion, and at the upper critical dimension, this is found to
vanish as $r_{\rm eff}(t) \rightarrow a/t^2$.
Due to the vanishing gap, there is no scale in the problem other than the time after the quench. In addition the lack
of inelastic scattering implies the system retains memory of the initial quench. Thus the solutions show
aging which is characterized by the emergence of universal scaling behavior.

The results  for the correlation ($G_K(k,t,t')=-i\langle \{\phi_k(t),\phi_{-k}(t')\} \rangle$) and the response
function ($G_R(k,t,t')=-i\langle \left[\phi_k(t),\phi_{-k}(t')\right]\rangle$) at two times $t$ and $t'$ after the quench,
as a function of the wavevector $k$ are
found to be,
\begin{subequations}
\begin{align}
G_K(k,t,t') & = \frac{1}{k^{2-2\theta_N}}\mathcal{G}_K(kt,kt'), \label{eq:GK-scaling-form}\\
G_R(k,t,t') & = \frac{1}{k} \left(\frac{t'}{t}\right)^{\theta_N} \mathcal{G}_R(kt,kt'), \label{eq:GR-scaling-form}
\end{align}
\end{subequations}
with $\theta_N=(N+2)\epsilon/(4 (N+8))$, and the scaling functions obeying,
$\mathcal{G}_K(x,y) \sim (xy)^{1-\theta_N}$ for $x,y\ll 1$ and $\mathcal{G}_K(x,y)\sim 1$ for $x,y\gg 1$, while
$\mathcal{G}_R(x,y) \sim x$ for $y\ll x\ll 1$~\cite{MitGam16}.

The exponent  $\theta_N$ appearing above is referred to as the initial slip exponent, and was first identified in quenches
in classical systems coupled to a thermalizing reservoir~\cite{Janssen1988,Huse89,Gambassi05}. The physical meaning of this
exponent is clarified as follows. Imagine modifying the quench by imposing a tiny magnetic field $\vec{\phi}\cdot\vec{h}$ in $H_i$.
At $t>0$, the quench involves time-evolving the system according to $H_f$ given above.
The difference now is that the initial state
has a small magnetization $M_0=\langle\vec{\phi}\rangle$ which acquires a time evolution.

Generic quenches into the gapped phase will cause the magnetization to decay exponentially in time (note that
our model does not conserve $\langle \vec{\phi}\rangle$). For a critical quench on the other hand, the susceptibility is enhanced.
Thus the initial response to $M_0$ is the growth of magnetization
with time $M(t) = M_0 t^{\theta_N}$.  Thus $\theta_N$ is the scaling dimension of the initial magnetic field.
Analogies can also be made with equilibrium phase transitions where spatial translation invariance is broken by a boundary, with
fields on the boundary acquiring a
different scaling dimension than the bulk fields~\cite{Diehlbook}. The $\theta_N$ exponent can also be associated with renormalization
of fields that now sit on the temporal boundary.

Although the exponent appearing in the growth of the initial magnetization is the same in the quantum and classical problem, there are important differences between the two systems. One is that for the quantum problem,
quasiparticles travel ballistically, whereas for the open classical system, quasiparticles travel diffusively.
Thus the quantum dynamics has a clear light-cone, and the precise temporal form of the response and correlation functions are not the same.

The difference between the finite-N and $N=\infty$ limit is that at the longest times, the former thermalizes, while the latter is
trapped in a nonequilibrium steady state which corresponds to the prethermal state
for the former. The larger is N, the more stable is the prethermal regime for the finite-N
model, and our perturbative RG scheme accurately captures this prethermal regime. Just like for the $1d$ problem, our RG also generates
dissipative terms in the effective action, with the strength of this self-generated dissipation providing
an estimate for the duration of the prethermal regime.

Eq.~\eqref{eq:GK-scaling-form} shows that the prethermal steady state has a bosonic occupation number which scales not
as $1/k^2$ as for the critical free theory, but sightly slower in momentum $1/k^{2-2\theta_N}$. This translates to a real space decay which is
faster in position for the critical interacting bosons in comparison to critical free bosons.

It is also useful to note that classically the total number of degrees of freedom in the system is $2N_d$, where $N_d$
of them are the ``positions'' $\phi_i$, and the other $N_d$ correspond to the ``momenta'' $\pi_j$. However, the quench dynamics
has only $N_d$ conserved quantities corresponding to fixing the commutation
relations between $\phi_i,\pi_j$. Thus strictly speaking the dynamics of the model is not integrable~\cite{Sondhi2013}.

\subsection{Entanglement dynamics}\label{hde}
An important question concerns the growth of entanglement entropy after a quench. There are
very few analytic and exact results on this topic in $d>1$ as the technical know-how developed for two
point correlation functions cannot be easily generalized
to the reduced density matrix. In an interacting system, Wick's theorem does not hold, therefore
to determine the density matrix one needs not only the two point correlation function but all higher moments, making
the task of calculating the entanglement statistics quite daunting.

Yet, the $N=\infty$ limit of field theories are effectively free as Hartree-Fock becomes exact. Moreover,
the results are not totally trivially related to a non-interacting theory as can be seen from the discussion of the
two point correlation above for the $O(N)$ model. Note that the non-trivial scaling exponent $\theta_N$ is
entirely due to interactions, and would vanish for non-interacting bosons.

Inspired by this observation, we exploited the quasi-free nature of the $\phi^4$ theory in the $N=\infty$ limit to
construct the reduced density matrix after the quench~\cite{Lemonik16}. We considered a real space entanglement cut in the form
of a hypercube of length $L$, constructed the two point correlations in that region, and from it constructed the reduced density matrix,
the last step following from the fact that Wicks' theorem holds~\cite{Eisler2009} when $N=\infty$.

In studying the critical quench described in the previous section, we found that the entanglement entropy (EE) grows linearly
in time at short times, saturating to a volume law at long times, where the latter is due to the finite density of
quasiparticles generated by the quench. This time evolution of the EE could be captured very well by
a semiclassical picture of ballistically propagating quasiparticles. Within this picture one assume particles
are emitted uniformly in all directions, but only the quasiparticles emitted in opposite directions are initially entangled
EPR pairs.
The EE at any given time was equated to the number of EPR pairs, where one member is inside the hypercube and the
other is outside. Two fitting parameters were used, one was the contribution to the EE by
an EPR pair, and the second was the EE at lattice time-scales. This semiclassical picture was found to fit the actual time evolution
of the EE very well indicating that the EE is not sensitive to the
subtle critical physics associated with aging~\cite{Lemonik16}.

In contrast the entanglement spectrum (ES) which are the eigenvalues of the reduced density matrix
show scaling which is sensitive to the initial slip exponent $\theta_N$.
If $\lambda$ are the eigenvalues of the correlation matrix in a hypercube of length $L$, then the eigenvalues of
the reduced density matrix $\omega$ are related to $\lambda$ as $2\lambda=\coth(\omega/2)$.
We found the following scaling form for all the $\lambda$ that constitute the "low-energy" part of the entanglement spectrum~\cite{Lemonik16}
\begin{eqnarray}
\lambda(L,t) = L^{-2\theta_N+1}W(t/L),
\end{eqnarray}
where $W$ is a scaling function.
Note that the scaling was obeyed by many eigenvalues corresponding to a substantial
fraction of the ES. When studying two point correlations,
there are only two independent correlators that depend on $\theta_N$. Thus for models where critical exponents are not apriori known,
the ES could be a good way to extract them as the same scaling form should be obeyed by a macroscopic number of quantities.

\section{Outlook}\label{conclu}
In the past it was thought that the main feature of nonequilibrium quantum systems was a high effective temperature, and consequently
such systems will show no quantum correlated states, with any correlations being fundamentally classical.
The recent hectic activity in the field shows this to be not always the case. There are
many body localized phases for example where quench dynamics can lead to steady states which are still coherent due to the suppression of
inelastic scattering. Even for clean systems, a long lived prethermal window can appear, which is characterized by a low
effective temperature, with nonequilibrium power-laws and scaling.

In the light of recent experiments in pump probe spectroscopy~\cite{Mitrano15}, an important open question is whether quenches can 
stabilize correlated states,
albeit in an intermediate time regime, different from those in thermal equilibrium. For example, due to the
existence of several competing superconducting orders in lattices, by simply tuning the quench amplitude
one may access superconducting orders of different symmetries~\cite{Dehghani17}.
Another important question of immediate interest is to study the long time limit after the quench where hydrodynamic modes are important.
One expects that the long thermalization times associated with conserved modes can lead to interesting time-dependence of observables.

\section{Acknowledgements}
The author is deeply indebted to A. Chiocchetta, A. Gambassi, T. Giamarchi, J. Lancaster, Y. Lemonik, J. Lux, A. Maraga, J. M\"{u}ller,
A. Rosch, L. Santos, M. Schir\'{o}, M. Tavora for valuable discussions and collaborations.
This work was supported by US National Science Foundation Grant No. DMR 1303177.

%\bibliography{quench}

\begin{thebibliography}{104}%
\makeatletter
\providecommand \@ifxundefined [1]{%
 \@ifx{#1\undefined}
}%
\providecommand \@ifnum [1]{%
 \ifnum #1\expandafter \@firstoftwo
 \else \expandafter \@secondoftwo
 \fi
}%
\providecommand \@ifx [1]{%
 \ifx #1\expandafter \@firstoftwo
 \else \expandafter \@secondoftwo
 \fi
}%
\providecommand \natexlab [1]{#1}%
\providecommand \enquote  [1]{``#1''}%
\providecommand \bibnamefont  [1]{#1}%
\providecommand \bibfnamefont [1]{#1}%
\providecommand \citenamefont [1]{#1}%
\providecommand \href@noop [0]{\@secondoftwo}%
\providecommand \href [0]{\begingroup \@sanitize@url \@href}%
\providecommand \@href[1]{\@@startlink{#1}\@@href}%
\providecommand \@@href[1]{\endgroup#1\@@endlink}%
\providecommand \@sanitize@url [0]{\catcode `\\12\catcode `\$12\catcode
  `\&12\catcode `\#12\catcode `\^12\catcode `\_12\catcode `\%12\relax}%
\providecommand \@@startlink[1]{}%
\providecommand \@@endlink[0]{}%
\providecommand \url  [0]{\begingroup\@sanitize@url \@url }%
\providecommand \@url [1]{\endgroup\@href {#1}{\urlprefix }}%
\providecommand \urlprefix  [0]{URL }%
\providecommand \Eprint [0]{\href }%
\providecommand \doibase [0]{http://dx.doi.org/}%
\providecommand \selectlanguage [0]{\@gobble}%
\providecommand \bibinfo  [0]{\@secondoftwo}%
\providecommand \bibfield  [0]{\@secondoftwo}%
\providecommand \translation [1]{[#1]}%
\providecommand \BibitemOpen [0]{}%
\providecommand \bibitemStop [0]{}%
\providecommand \bibitemNoStop [0]{.\EOS\space}%
\providecommand \EOS [0]{\spacefactor3000\relax}%
\providecommand \BibitemShut  [1]{\csname bibitem#1\endcsname}%
\let\auto@bib@innerbib\@empty
%</preamble>
\bibitem [{\citenamefont {Calabrese}\ and\ \citenamefont
  {Cardy}(2005)}]{Calabrese05b}%
  \BibitemOpen
  \bibfield  {author} {\bibinfo {author} {\bibfnamefont {P.}~\bibnamefont
  {Calabrese}}\ and\ \bibinfo {author} {\bibfnamefont {J.}~\bibnamefont
  {Cardy}},\ }\href {http://stacks.iop.org/1742-5468/2005/i=04/a=P04010}
  {\bibfield  {journal} {\bibinfo  {journal} {Journal of Statistical Mechanics:
  Theory and Experiment}\ ,\ \bibinfo {pages} {P04010}} (\bibinfo {year}
  {2005})}\BibitemShut {NoStop}%
\bibitem [{\citenamefont {Calabrese}\ and\ \citenamefont
  {Cardy}(2006)}]{Calabrese2006}%
  \BibitemOpen
  \bibfield  {author} {\bibinfo {author} {\bibfnamefont {P.}~\bibnamefont
  {Calabrese}}\ and\ \bibinfo {author} {\bibfnamefont {J.}~\bibnamefont
  {Cardy}},\ }\href {\doibase 10.1103/PhysRevLett.96.136801} {\bibfield
  {journal} {\bibinfo  {journal} {Phys. Rev. Lett.}\ }\textbf {\bibinfo
  {volume} {96}},\ \bibinfo {pages} {136801} (\bibinfo {year}
  {2006})}\BibitemShut {NoStop}%
\bibitem [{\citenamefont {Bloch}\ \emph {et~al.}(2008)\citenamefont {Bloch},
  \citenamefont {Dalibard},\ and\ \citenamefont {Zwerger}}]{Bloch08}%
  \BibitemOpen
  \bibfield  {author} {\bibinfo {author} {\bibfnamefont {I.}~\bibnamefont
  {Bloch}}, \bibinfo {author} {\bibfnamefont {J.}~\bibnamefont {Dalibard}}, \
  and\ \bibinfo {author} {\bibfnamefont {W.}~\bibnamefont {Zwerger}},\ }\href
  {\doibase 10.1103/RevModPhys.80.885} {\bibfield  {journal} {\bibinfo
  {journal} {Rev. Mod. Phys.}\ }\textbf {\bibinfo {volume} {80}},\ \bibinfo
  {pages} {885} (\bibinfo {year} {2008})}\BibitemShut {NoStop}%
\bibitem [{\citenamefont {Cazalilla}\ \emph {et~al.}(2011)\citenamefont
  {Cazalilla}, \citenamefont {Citro}, \citenamefont {Giamarchi}, \citenamefont
  {Orignac},\ and\ \citenamefont {Rigol}}]{Cazalillarev11}%
  \BibitemOpen
  \bibfield  {author} {\bibinfo {author} {\bibfnamefont {M.~A.}\ \bibnamefont
  {Cazalilla}}, \bibinfo {author} {\bibfnamefont {R.}~\bibnamefont {Citro}},
  \bibinfo {author} {\bibfnamefont {T.}~\bibnamefont {Giamarchi}}, \bibinfo
  {author} {\bibfnamefont {E.}~\bibnamefont {Orignac}}, \ and\ \bibinfo
  {author} {\bibfnamefont {M.}~\bibnamefont {Rigol}},\ }\href {\doibase
  10.1103/RevModPhys.83.1405} {\bibfield  {journal} {\bibinfo  {journal} {Rev.
  Mod. Phys.}\ }\textbf {\bibinfo {volume} {83}},\ \bibinfo {pages} {1405}
  (\bibinfo {year} {2011})}\BibitemShut {NoStop}%
\bibitem [{\citenamefont {Kinoshita}\ \emph {et~al.}(2006)\citenamefont
  {Kinoshita}, \citenamefont {Wenger},\ and\ \citenamefont {Weiss}}]{Weiss06}%
  \BibitemOpen
  \bibfield  {author} {\bibinfo {author} {\bibfnamefont {T.}~\bibnamefont
  {Kinoshita}}, \bibinfo {author} {\bibfnamefont {T.}~\bibnamefont {Wenger}}, \
  and\ \bibinfo {author} {\bibfnamefont {D.~S.}\ \bibnamefont {Weiss}},\
  }\href@noop {} {\bibfield  {journal} {\bibinfo  {journal} {Nature (London)}\
  }\textbf {\bibinfo {volume} {440}},\ \bibinfo {pages} {900} (\bibinfo {year}
  {2006})}\BibitemShut {NoStop}%
\bibitem [{\citenamefont {Gring}\ \emph {et~al.}(2012)\citenamefont {Gring},
  \citenamefont {Kuhnert}, \citenamefont {Langen}, \citenamefont {Kitagawa},
  \citenamefont {Rauer}, \citenamefont {Schreitl}, \citenamefont {Mazets},
  \citenamefont {Smith}, \citenamefont {Demler},\ and\ \citenamefont
  {Schmiedmayer}}]{Gring12}%
  \BibitemOpen
  \bibfield  {author} {\bibinfo {author} {\bibfnamefont {M.}~\bibnamefont
  {Gring}}, \bibinfo {author} {\bibfnamefont {M.}~\bibnamefont {Kuhnert}},
  \bibinfo {author} {\bibfnamefont {T.}~\bibnamefont {Langen}}, \bibinfo
  {author} {\bibfnamefont {T.}~\bibnamefont {Kitagawa}}, \bibinfo {author}
  {\bibfnamefont {B.}~\bibnamefont {Rauer}}, \bibinfo {author} {\bibfnamefont
  {M.}~\bibnamefont {Schreitl}}, \bibinfo {author} {\bibfnamefont
  {I.}~\bibnamefont {Mazets}}, \bibinfo {author} {\bibfnamefont {D.~A.}\
  \bibnamefont {Smith}}, \bibinfo {author} {\bibfnamefont {E.}~\bibnamefont
  {Demler}}, \ and\ \bibinfo {author} {\bibfnamefont {J.}~\bibnamefont
  {Schmiedmayer}},\ }\href@noop {} {\bibfield  {journal} {\bibinfo  {journal}
  {Science}\ }\textbf {\bibinfo {volume} {337}},\ \bibinfo {pages} {1318}
  (\bibinfo {year} {2012})}\BibitemShut {NoStop}%
\bibitem [{\citenamefont {Trotzky}\ \emph {et~al.}(2012)\citenamefont
  {Trotzky}, \citenamefont {Chen}, \citenamefont {Flesch}, \citenamefont
  {McCulloch}, \citenamefont {Schollw\"ock}, \citenamefont {Eisert},\ and\
  \citenamefont {Bloch}}]{Trotzky12}%
  \BibitemOpen
  \bibfield  {author} {\bibinfo {author} {\bibfnamefont {S.}~\bibnamefont
  {Trotzky}}, \bibinfo {author} {\bibfnamefont {Y.-A.}\ \bibnamefont {Chen}},
  \bibinfo {author} {\bibfnamefont {A.}~\bibnamefont {Flesch}}, \bibinfo
  {author} {\bibfnamefont {I.}~\bibnamefont {McCulloch}}, \bibinfo {author}
  {\bibfnamefont {U.}~\bibnamefont {Schollw\"ock}}, \bibinfo {author}
  {\bibfnamefont {J.}~\bibnamefont {Eisert}}, \ and\ \bibinfo {author}
  {\bibfnamefont {I.}~\bibnamefont {Bloch}},\ }\href@noop {} {\bibfield
  {journal} {\bibinfo  {journal} {Nature Physics}\ }\textbf {\bibinfo {volume}
  {8}},\ \bibinfo {pages} {325} (\bibinfo {year} {2012})}\BibitemShut {NoStop}%
\bibitem [{\citenamefont {Schreiber}\ \emph {et~al.}(2015)\citenamefont
  {Schreiber}, \citenamefont {Hodgman}, \citenamefont {Bordia}, \citenamefont
  {Lüschen}, \citenamefont {Fischer}, \citenamefont {Vosk}, \citenamefont
  {Altman}, \citenamefont {Schneider},\ and\ \citenamefont {Bloch}}]{Bloch15}%
  \BibitemOpen
  \bibfield  {author} {\bibinfo {author} {\bibfnamefont {M.}~\bibnamefont
  {Schreiber}}, \bibinfo {author} {\bibfnamefont {S.~S.}\ \bibnamefont
  {Hodgman}}, \bibinfo {author} {\bibfnamefont {P.}~\bibnamefont {Bordia}},
  \bibinfo {author} {\bibfnamefont {H.~P.}\ \bibnamefont {Lüschen}}, \bibinfo
  {author} {\bibfnamefont {M.~H.}\ \bibnamefont {Fischer}}, \bibinfo {author}
  {\bibfnamefont {R.}~\bibnamefont {Vosk}}, \bibinfo {author} {\bibfnamefont
  {E.}~\bibnamefont {Altman}}, \bibinfo {author} {\bibfnamefont
  {U.}~\bibnamefont {Schneider}}, \ and\ \bibinfo {author} {\bibfnamefont
  {I.}~\bibnamefont {Bloch}},\ }\href {\doibase 10.1126/science.aaa7432}
  {\bibfield  {journal} {\bibinfo  {journal} {Science}\ }\textbf {\bibinfo
  {volume} {349}},\ \bibinfo {pages} {842} (\bibinfo {year}
  {2015})}\BibitemShut {NoStop}%
\bibitem [{\citenamefont {Deutsch}(1991)}]{Deutsch91}%
  \BibitemOpen
  \bibfield  {author} {\bibinfo {author} {\bibfnamefont {J.~M.}\ \bibnamefont
  {Deutsch}},\ }\href {\doibase 10.1103/PhysRevA.43.2046} {\bibfield  {journal}
  {\bibinfo  {journal} {Phys. Rev. A}\ }\textbf {\bibinfo {volume} {43}},\
  \bibinfo {pages} {2046} (\bibinfo {year} {1991})}\BibitemShut {NoStop}%
\bibitem [{\citenamefont {Srednicki}(1994)}]{Srednicki94}%
  \BibitemOpen
  \bibfield  {author} {\bibinfo {author} {\bibfnamefont {M.}~\bibnamefont
  {Srednicki}},\ }\href {\doibase 10.1103/PhysRevE.50.888} {\bibfield
  {journal} {\bibinfo  {journal} {Phys. Rev. E}\ }\textbf {\bibinfo {volume}
  {50}},\ \bibinfo {pages} {888} (\bibinfo {year} {1994})}\BibitemShut
  {NoStop}%
\bibitem [{\citenamefont {Rigol}\ \emph {et~al.}(2008)\citenamefont {Rigol},
  \citenamefont {Dunjko}, \citenamefont {Yurovsky},\ and\ \citenamefont
  {Olshanii}}]{Rigol08}%
  \BibitemOpen
  \bibfield  {author} {\bibinfo {author} {\bibfnamefont {M.}~\bibnamefont
  {Rigol}}, \bibinfo {author} {\bibfnamefont {V.}~\bibnamefont {Dunjko}},
  \bibinfo {author} {\bibfnamefont {V.}~\bibnamefont {Yurovsky}}, \ and\
  \bibinfo {author} {\bibfnamefont {M.}~\bibnamefont {Olshanii}},\ }\href@noop
  {} {\bibfield  {journal} {\bibinfo  {journal} {Nature}\ }\textbf {\bibinfo
  {volume} {452}},\ \bibinfo {pages} {854} (\bibinfo {year}
  {2008})}\BibitemShut {NoStop}%
\bibitem [{\citenamefont {Iyer}\ and\ \citenamefont {Andrei}(2012)}]{Andrei12}%
  \BibitemOpen
  \bibfield  {author} {\bibinfo {author} {\bibfnamefont {D.}~\bibnamefont
  {Iyer}}\ and\ \bibinfo {author} {\bibfnamefont {N.}~\bibnamefont {Andrei}},\
  }\href {\doibase 10.1103/PhysRevLett.109.115304} {\bibfield  {journal}
  {\bibinfo  {journal} {Phys. Rev. Lett.}\ }\textbf {\bibinfo {volume} {109}},\
  \bibinfo {pages} {115304} (\bibinfo {year} {2012})}\BibitemShut {NoStop}%
\bibitem [{\citenamefont {Iyer}\ \emph {et~al.}(2013)\citenamefont {Iyer},
  \citenamefont {Guan},\ and\ \citenamefont {Andrei}}]{Iyer13}%
  \BibitemOpen
  \bibfield  {author} {\bibinfo {author} {\bibfnamefont {D.}~\bibnamefont
  {Iyer}}, \bibinfo {author} {\bibfnamefont {H.}~\bibnamefont {Guan}}, \ and\
  \bibinfo {author} {\bibfnamefont {N.}~\bibnamefont {Andrei}},\ }\href
  {\doibase 10.1103/PhysRevA.87.053628} {\bibfield  {journal} {\bibinfo
  {journal} {Phys. Rev. A}\ }\textbf {\bibinfo {volume} {87}},\ \bibinfo
  {pages} {053628} (\bibinfo {year} {2013})}\BibitemShut {NoStop}%
\bibitem [{\citenamefont {Mussardo}(2013)}]{Mussardo13}%
  \BibitemOpen
  \bibfield  {author} {\bibinfo {author} {\bibfnamefont {G.}~\bibnamefont
  {Mussardo}},\ }\href {\doibase 10.1103/PhysRevLett.111.100401} {\bibfield
  {journal} {\bibinfo  {journal} {Phys. Rev. Lett.}\ }\textbf {\bibinfo
  {volume} {111}},\ \bibinfo {pages} {100401} (\bibinfo {year}
  {2013})}\BibitemShut {NoStop}%
\bibitem [{\citenamefont {van~den Berg}\ \emph {et~al.}(2016)\citenamefont
  {van~den Berg}, \citenamefont {Wouters}, \citenamefont {Eli\"ens},
  \citenamefont {De~Nardis}, \citenamefont {Konik},\ and\ \citenamefont
  {Caux}}]{Wouter16}%
  \BibitemOpen
  \bibfield  {author} {\bibinfo {author} {\bibfnamefont {R.}~\bibnamefont
  {van~den Berg}}, \bibinfo {author} {\bibfnamefont {B.}~\bibnamefont
  {Wouters}}, \bibinfo {author} {\bibfnamefont {S.}~\bibnamefont {Eli\"ens}},
  \bibinfo {author} {\bibfnamefont {J.}~\bibnamefont {De~Nardis}}, \bibinfo
  {author} {\bibfnamefont {R.~M.}\ \bibnamefont {Konik}}, \ and\ \bibinfo
  {author} {\bibfnamefont {J.-S.}\ \bibnamefont {Caux}},\ }\href {\doibase
  10.1103/PhysRevLett.116.225302} {\bibfield  {journal} {\bibinfo  {journal}
  {Phys. Rev. Lett.}\ }\textbf {\bibinfo {volume} {116}},\ \bibinfo {pages}
  {225302} (\bibinfo {year} {2016})}\BibitemShut {NoStop}%
\bibitem [{\citenamefont {Yuzbashyan}\ and\ \citenamefont
  {Shastry}(2013)}]{Yuzbashyan11}%
  \BibitemOpen
  \bibfield  {author} {\bibinfo {author} {\bibfnamefont {E.~A.}\ \bibnamefont
  {Yuzbashyan}}\ and\ \bibinfo {author} {\bibfnamefont {B.~S.}\ \bibnamefont
  {Shastry}},\ }\href@noop {} {\bibfield  {journal} {\bibinfo  {journal} {J.
  Stat. Phys.}\ }\textbf {\bibinfo {volume} {150}},\ \bibinfo {pages} {704}
  (\bibinfo {year} {2013})}\BibitemShut {NoStop}%
\bibitem [{\citenamefont {Kolmogorov}(1954)}]{KAM}%
  \BibitemOpen
  \bibfield  {author} {\bibinfo {author} {\bibfnamefont {A.~N.}\ \bibnamefont
  {Kolmogorov}},\ }\href@noop {} {\bibfield  {journal} {\bibinfo  {journal}
  {Dokl. Akad. Nauk SSSR}\ }\textbf {\bibinfo {volume} {98}},\ \bibinfo {pages}
  {527} (\bibinfo {year} {1954})}\BibitemShut {NoStop}%
\bibitem [{\citenamefont {Brandino}\ \emph {et~al.}(2015)\citenamefont
  {Brandino}, \citenamefont {Caux},\ and\ \citenamefont {Konik}}]{Konik15}%
  \BibitemOpen
  \bibfield  {author} {\bibinfo {author} {\bibfnamefont {G.~P.}\ \bibnamefont
  {Brandino}}, \bibinfo {author} {\bibfnamefont {J.-S.}\ \bibnamefont {Caux}},
  \ and\ \bibinfo {author} {\bibfnamefont {R.~M.}\ \bibnamefont {Konik}},\
  }\href {\doibase 10.1103/PhysRevX.5.041043} {\bibfield  {journal} {\bibinfo
  {journal} {Phys. Rev. X}\ }\textbf {\bibinfo {volume} {5}},\ \bibinfo {pages}
  {041043} (\bibinfo {year} {2015})}\BibitemShut {NoStop}%
\bibitem [{\citenamefont {Anderson}(1958)}]{Anderson58}%
  \BibitemOpen
  \bibfield  {author} {\bibinfo {author} {\bibfnamefont {P.~W.}\ \bibnamefont
  {Anderson}},\ }\href {\doibase 10.1103/PhysRev.109.1492} {\bibfield
  {journal} {\bibinfo  {journal} {Phys. Rev.}\ }\textbf {\bibinfo {volume}
  {109}},\ \bibinfo {pages} {1492} (\bibinfo {year} {1958})}\BibitemShut
  {NoStop}%
\bibitem [{\citenamefont {Basko}\ \emph {et~al.}(2006)\citenamefont {Basko},
  \citenamefont {Aleiner},\ and\ \citenamefont {Altshuler}}]{Basko06}%
  \BibitemOpen
  \bibfield  {author} {\bibinfo {author} {\bibfnamefont {D.}~\bibnamefont
  {Basko}}, \bibinfo {author} {\bibfnamefont {I.}~\bibnamefont {Aleiner}}, \
  and\ \bibinfo {author} {\bibfnamefont {B.}~\bibnamefont {Altshuler}},\ }\href
  {\doibase http://dx.doi.org/10.1016/j.aop.2005.11.014} {\bibfield  {journal}
  {\bibinfo  {journal} {Annals of Physics}\ }\textbf {\bibinfo {volume}
  {321}},\ \bibinfo {pages} {1126 } (\bibinfo {year} {2006})}\BibitemShut
  {NoStop}%
\bibitem [{\citenamefont {Oganesyan}\ and\ \citenamefont
  {Huse}(2007)}]{Oganesyan07}%
  \BibitemOpen
  \bibfield  {author} {\bibinfo {author} {\bibfnamefont {V.}~\bibnamefont
  {Oganesyan}}\ and\ \bibinfo {author} {\bibfnamefont {D.~A.}\ \bibnamefont
  {Huse}},\ }\href {\doibase 10.1103/PhysRevB.75.155111} {\bibfield  {journal}
  {\bibinfo  {journal} {Phys. Rev. B}\ }\textbf {\bibinfo {volume} {75}},\
  \bibinfo {pages} {155111} (\bibinfo {year} {2007})}\BibitemShut {NoStop}%
\bibitem [{\citenamefont {Nandkishore}\ and\ \citenamefont
  {Huse}(2015)}]{Huse15}%
  \BibitemOpen
  \bibfield  {author} {\bibinfo {author} {\bibfnamefont {R.}~\bibnamefont
  {Nandkishore}}\ and\ \bibinfo {author} {\bibfnamefont {D.~A.}\ \bibnamefont
  {Huse}},\ }\href {\doibase 10.1146/annurev-conmatphys-031214-014726}
  {\bibfield  {journal} {\bibinfo  {journal} {Annual Review of Condensed Matter
  Physics}\ }\textbf {\bibinfo {volume} {6}},\ \bibinfo {pages} {15} (\bibinfo
  {year} {2015})}\BibitemShut {NoStop}%
\bibitem [{\citenamefont {Vidal}\ \emph {et~al.}(2003)\citenamefont {Vidal},
  \citenamefont {Latorre}, \citenamefont {Rico},\ and\ \citenamefont
  {Kitaev}}]{Vidal03}%
  \BibitemOpen
  \bibfield  {author} {\bibinfo {author} {\bibfnamefont {G.}~\bibnamefont
  {Vidal}}, \bibinfo {author} {\bibfnamefont {J.~I.}\ \bibnamefont {Latorre}},
  \bibinfo {author} {\bibfnamefont {E.}~\bibnamefont {Rico}}, \ and\ \bibinfo
  {author} {\bibfnamefont {A.}~\bibnamefont {Kitaev}},\ }\href {\doibase
  10.1103/PhysRevLett.90.227902} {\bibfield  {journal} {\bibinfo  {journal}
  {Phys. Rev. Lett.}\ }\textbf {\bibinfo {volume} {90}},\ \bibinfo {pages}
  {227902} (\bibinfo {year} {2003})}\BibitemShut {NoStop}%
\bibitem [{\citenamefont {Kitaev}\ and\ \citenamefont
  {Preskill}(2006)}]{Preskill06}%
  \BibitemOpen
  \bibfield  {author} {\bibinfo {author} {\bibfnamefont {A.}~\bibnamefont
  {Kitaev}}\ and\ \bibinfo {author} {\bibfnamefont {J.}~\bibnamefont
  {Preskill}},\ }\href {\doibase 10.1103/PhysRevLett.96.110404} {\bibfield
  {journal} {\bibinfo  {journal} {Phys. Rev. Lett.}\ }\textbf {\bibinfo
  {volume} {96}},\ \bibinfo {pages} {110404} (\bibinfo {year}
  {2006})}\BibitemShut {NoStop}%
\bibitem [{\citenamefont {Levin}\ and\ \citenamefont {Wen}(2006)}]{Levin06}%
  \BibitemOpen
  \bibfield  {author} {\bibinfo {author} {\bibfnamefont {M.}~\bibnamefont
  {Levin}}\ and\ \bibinfo {author} {\bibfnamefont {X.-G.}\ \bibnamefont
  {Wen}},\ }\href {\doibase 10.1103/PhysRevLett.96.110405} {\bibfield
  {journal} {\bibinfo  {journal} {Phys. Rev. Lett.}\ }\textbf {\bibinfo
  {volume} {96}},\ \bibinfo {pages} {110405} (\bibinfo {year}
  {2006})}\BibitemShut {NoStop}%
\bibitem [{\citenamefont {Li}\ and\ \citenamefont {Haldane}(2008)}]{Haldane08}%
  \BibitemOpen
  \bibfield  {author} {\bibinfo {author} {\bibfnamefont {H.}~\bibnamefont
  {Li}}\ and\ \bibinfo {author} {\bibfnamefont {F.~D.~M.}\ \bibnamefont
  {Haldane}},\ }\href {\doibase 10.1103/PhysRevLett.101.010504} {\bibfield
  {journal} {\bibinfo  {journal} {Phys. Rev. Lett.}\ }\textbf {\bibinfo
  {volume} {101}},\ \bibinfo {pages} {010504} (\bibinfo {year}
  {2008})}\BibitemShut {NoStop}%
\bibitem [{\citenamefont {Casini}\ and\ \citenamefont
  {Huerta}(2009)}]{Casini09}%
  \BibitemOpen
  \bibfield  {author} {\bibinfo {author} {\bibfnamefont {H.}~\bibnamefont
  {Casini}}\ and\ \bibinfo {author} {\bibfnamefont {M.}~\bibnamefont
  {Huerta}},\ }\href {http://stacks.iop.org/1751-8121/42/i=50/a=504007}
  {\bibfield  {journal} {\bibinfo  {journal} {Journal of Physics A:
  Mathematical and Theoretical}\ }\textbf {\bibinfo {volume} {42}},\ \bibinfo
  {pages} {504007} (\bibinfo {year} {2009})}\BibitemShut {NoStop}%
\bibitem [{\citenamefont {Eisert}\ \emph {et~al.}(2010)\citenamefont {Eisert},
  \citenamefont {Cramer},\ and\ \citenamefont {Plenio}}]{Eisert10}%
  \BibitemOpen
  \bibfield  {author} {\bibinfo {author} {\bibfnamefont {J.}~\bibnamefont
  {Eisert}}, \bibinfo {author} {\bibfnamefont {M.}~\bibnamefont {Cramer}}, \
  and\ \bibinfo {author} {\bibfnamefont {M.~B.}\ \bibnamefont {Plenio}},\
  }\href {\doibase 10.1103/RevModPhys.82.277} {\bibfield  {journal} {\bibinfo
  {journal} {Rev. Mod. Phys.}\ }\textbf {\bibinfo {volume} {82}},\ \bibinfo
  {pages} {277} (\bibinfo {year} {2010})}\BibitemShut {NoStop}%
\bibitem [{\citenamefont {Calabrese}\ and\ \citenamefont
  {Cardy}(2007{\natexlab{a}})}]{Calabrese2007}%
  \BibitemOpen
  \bibfield  {author} {\bibinfo {author} {\bibfnamefont {P.}~\bibnamefont
  {Calabrese}}\ and\ \bibinfo {author} {\bibfnamefont {J.}~\bibnamefont
  {Cardy}},\ }\href {http://stacks.iop.org/1742-5468/2007/i=06/a=P06008}
  {\bibfield  {journal} {\bibinfo  {journal} {J. Stat. Mech.}\ ,\ \bibinfo
  {pages} {P06008}} (\bibinfo {year} {2007}{\natexlab{a}})}\BibitemShut
  {NoStop}%
\bibitem [{\citenamefont {Calabrese}\ and\ \citenamefont
  {Cardy}(2007{\natexlab{b}})}]{Calabrese07b}%
  \BibitemOpen
  \bibfield  {author} {\bibinfo {author} {\bibfnamefont {P.}~\bibnamefont
  {Calabrese}}\ and\ \bibinfo {author} {\bibfnamefont {J.}~\bibnamefont
  {Cardy}},\ }\href {http://stacks.iop.org/1742-5468/2007/i=10/a=P10004}
  {\bibfield  {journal} {\bibinfo  {journal} {Journal of Statistical Mechanics:
  Theory and Experiment}\ ,\ \bibinfo {pages} {P10004}} (\bibinfo {year}
  {2007}{\natexlab{b}})}\BibitemShut {NoStop}%
\bibitem [{\citenamefont {Hartman}\ and\ \citenamefont
  {Maldacena}(2013)}]{Hartman2013}%
  \BibitemOpen
  \bibfield  {author} {\bibinfo {author} {\bibfnamefont {T.}~\bibnamefont
  {Hartman}}\ and\ \bibinfo {author} {\bibfnamefont {J.}~\bibnamefont
  {Maldacena}},\ }\href {\doibase 10.1007/JHEP05(2013)014} {\bibfield
  {journal} {\bibinfo  {journal} {Journal of High Energy Physics}\ }\textbf
  {\bibinfo {volume} {2013}},\ \bibinfo {pages} {14} (\bibinfo {year}
  {2013})}\BibitemShut {NoStop}%
\bibitem [{\citenamefont {Igl\'oi}\ \emph {et~al.}(2012)\citenamefont
  {Igl\'oi}, \citenamefont {Szatm\'ari},\ and\ \citenamefont {Lin}}]{Lin12}%
  \BibitemOpen
  \bibfield  {author} {\bibinfo {author} {\bibfnamefont {F.}~\bibnamefont
  {Igl\'oi}}, \bibinfo {author} {\bibfnamefont {Z.}~\bibnamefont {Szatm\'ari}},
  \ and\ \bibinfo {author} {\bibfnamefont {Y.-C.}\ \bibnamefont {Lin}},\ }\href
  {\doibase 10.1103/PhysRevB.85.094417} {\bibfield  {journal} {\bibinfo
  {journal} {Phys. Rev. B}\ }\textbf {\bibinfo {volume} {85}},\ \bibinfo
  {pages} {094417} (\bibinfo {year} {2012})}\BibitemShut {NoStop}%
\bibitem [{\citenamefont {Kim}\ and\ \citenamefont {Huse}(2013)}]{Huse13}%
  \BibitemOpen
  \bibfield  {author} {\bibinfo {author} {\bibfnamefont {H.}~\bibnamefont
  {Kim}}\ and\ \bibinfo {author} {\bibfnamefont {D.~A.}\ \bibnamefont {Huse}},\
  }\href {\doibase 10.1103/PhysRevLett.111.127205} {\bibfield  {journal}
  {\bibinfo  {journal} {Phys. Rev. Lett.}\ }\textbf {\bibinfo {volume} {111}},\
  \bibinfo {pages} {127205} (\bibinfo {year} {2013})}\BibitemShut {NoStop}%
\bibitem [{\citenamefont {Schollwock}(2011)}]{TDMRGrev11}%
  \BibitemOpen
  \bibfield  {author} {\bibinfo {author} {\bibfnamefont {U.}~\bibnamefont
  {Schollwock}},\ }\href {\doibase http://dx.doi.org/10.1016/j.aop.2010.09.012}
  {\bibfield  {journal} {\bibinfo  {journal} {Annals of Physics}\ }\textbf
  {\bibinfo {volume} {326}},\ \bibinfo {pages} {96 } (\bibinfo {year}
  {2011})},\ \bibinfo {note} {january 2011 Special Issue}\BibitemShut {NoStop}%
\bibitem [{\citenamefont {Haegeman}\ \emph {et~al.}(2011)\citenamefont
  {Haegeman}, \citenamefont {Cirac}, \citenamefont {Osborne}, \citenamefont
  {Pi\ifmmode~\check{z}\else \v{z}\fi{}orn}, \citenamefont {Verschelde},\ and\
  \citenamefont {Verstraete}}]{Cirac11}%
  \BibitemOpen
  \bibfield  {author} {\bibinfo {author} {\bibfnamefont {J.}~\bibnamefont
  {Haegeman}}, \bibinfo {author} {\bibfnamefont {J.~I.}\ \bibnamefont {Cirac}},
  \bibinfo {author} {\bibfnamefont {T.~J.}\ \bibnamefont {Osborne}}, \bibinfo
  {author} {\bibfnamefont {I.}~\bibnamefont {Pi\ifmmode~\check{z}\else
  \v{z}\fi{}orn}}, \bibinfo {author} {\bibfnamefont {H.}~\bibnamefont
  {Verschelde}}, \ and\ \bibinfo {author} {\bibfnamefont {F.}~\bibnamefont
  {Verstraete}},\ }\href {\doibase 10.1103/PhysRevLett.107.070601} {\bibfield
  {journal} {\bibinfo  {journal} {Phys. Rev. Lett.}\ }\textbf {\bibinfo
  {volume} {107}},\ \bibinfo {pages} {070601} (\bibinfo {year}
  {2011})}\BibitemShut {NoStop}%
\bibitem [{\citenamefont {Leviatan}\ \emph {et~al.}(shed)\citenamefont
  {Leviatan}, \citenamefont {Pollmann}, \citenamefont {Bardarson},\ and\
  \citenamefont {Altman}}]{Altman17}%
  \BibitemOpen
  \bibfield  {author} {\bibinfo {author} {\bibfnamefont {E.}~\bibnamefont
  {Leviatan}}, \bibinfo {author} {\bibfnamefont {F.}~\bibnamefont {Pollmann}},
  \bibinfo {author} {\bibfnamefont {J.~H.}\ \bibnamefont {Bardarson}}, \ and\
  \bibinfo {author} {\bibfnamefont {E.}~\bibnamefont {Altman}},\ }\href@noop {}
  {\bibfield  {journal} {\bibinfo  {journal} {arXiv:1702.08894}\ } (\bibinfo
  {year} {unpublished})}\BibitemShut {NoStop}%
\bibitem [{\citenamefont {Giamarchi}(2004)}]{Giamarchibook}%
  \BibitemOpen
  \bibfield  {author} {\bibinfo {author} {\bibfnamefont {T.}~\bibnamefont
  {Giamarchi}},\ }\href@noop {} {\bibfield  {journal} {\bibinfo  {journal}
  {{\sl Quantum Physics in One Dimension}, Oxford University Press, Oxford}\ }
  (\bibinfo {year} {2004})}\BibitemShut {NoStop}%
\bibitem [{\citenamefont {Berges}\ \emph {et~al.}(2004)\citenamefont {Berges},
  \citenamefont {Bors\'anyi},\ and\ \citenamefont {Wetterich}}]{Berges04}%
  \BibitemOpen
  \bibfield  {author} {\bibinfo {author} {\bibfnamefont {J.}~\bibnamefont
  {Berges}}, \bibinfo {author} {\bibfnamefont {S.}~\bibnamefont {Bors\'anyi}},
  \ and\ \bibinfo {author} {\bibfnamefont {C.}~\bibnamefont {Wetterich}},\
  }\href {\doibase 10.1103/PhysRevLett.93.142002} {\bibfield  {journal}
  {\bibinfo  {journal} {Phys. Rev. Lett.}\ }\textbf {\bibinfo {volume} {93}},\
  \bibinfo {pages} {142002} (\bibinfo {year} {2004})}\BibitemShut {NoStop}%
\bibitem [{\citenamefont {Moeckel}\ and\ \citenamefont
  {Kehrein}(2008)}]{Kehrein08}%
  \BibitemOpen
  \bibfield  {author} {\bibinfo {author} {\bibfnamefont {M.}~\bibnamefont
  {Moeckel}}\ and\ \bibinfo {author} {\bibfnamefont {S.}~\bibnamefont
  {Kehrein}},\ }\href {\doibase 10.1103/PhysRevLett.100.175702} {\bibfield
  {journal} {\bibinfo  {journal} {Phys. Rev. Lett.}\ }\textbf {\bibinfo
  {volume} {100}},\ \bibinfo {pages} {175702} (\bibinfo {year}
  {2008})}\BibitemShut {NoStop}%
\bibitem [{\citenamefont {Kollar}\ \emph {et~al.}(2011)\citenamefont {Kollar},
  \citenamefont {Wolf},\ and\ \citenamefont {Eckstein}}]{Kollar11}%
  \BibitemOpen
  \bibfield  {author} {\bibinfo {author} {\bibfnamefont {M.}~\bibnamefont
  {Kollar}}, \bibinfo {author} {\bibfnamefont {F.~A.}\ \bibnamefont {Wolf}}, \
  and\ \bibinfo {author} {\bibfnamefont {M.}~\bibnamefont {Eckstein}},\ }\href
  {\doibase 10.1103/PhysRevB.84.054304} {\bibfield  {journal} {\bibinfo
  {journal} {Phys. Rev. B}\ }\textbf {\bibinfo {volume} {84}},\ \bibinfo
  {pages} {054304} (\bibinfo {year} {2011})}\BibitemShut {NoStop}%
\bibitem [{\citenamefont {Marcuzzi}\ \emph {et~al.}(2013)\citenamefont
  {Marcuzzi}, \citenamefont {Marino}, \citenamefont {Gambassi},\ and\
  \citenamefont {Silva}}]{Marcuzzi13}%
  \BibitemOpen
  \bibfield  {author} {\bibinfo {author} {\bibfnamefont {M.}~\bibnamefont
  {Marcuzzi}}, \bibinfo {author} {\bibfnamefont {J.}~\bibnamefont {Marino}},
  \bibinfo {author} {\bibfnamefont {A.}~\bibnamefont {Gambassi}}, \ and\
  \bibinfo {author} {\bibfnamefont {A.}~\bibnamefont {Silva}},\ }\href
  {\doibase 10.1103/PhysRevLett.111.197203} {\bibfield  {journal} {\bibinfo
  {journal} {Phys. Rev. Lett.}\ }\textbf {\bibinfo {volume} {111}},\ \bibinfo
  {pages} {197203} (\bibinfo {year} {2013})}\BibitemShut {NoStop}%
\bibitem [{\citenamefont {Antal}\ \emph {et~al.}(1999)\citenamefont {Antal},
  \citenamefont {R\'acz}, \citenamefont {R\'akos},\ and\ \citenamefont
  {Sch\"utz}}]{Antal99}%
  \BibitemOpen
  \bibfield  {author} {\bibinfo {author} {\bibfnamefont {T.}~\bibnamefont
  {Antal}}, \bibinfo {author} {\bibfnamefont {Z.}~\bibnamefont {R\'acz}},
  \bibinfo {author} {\bibfnamefont {A.}~\bibnamefont {R\'akos}}, \ and\
  \bibinfo {author} {\bibfnamefont {G.~M.}\ \bibnamefont {Sch\"utz}},\ }\href
  {\doibase 10.1103/PhysRevE.59.4912} {\bibfield  {journal} {\bibinfo
  {journal} {Phys. Rev. E}\ }\textbf {\bibinfo {volume} {59}},\ \bibinfo
  {pages} {4912} (\bibinfo {year} {1999})}\BibitemShut {NoStop}%
\bibitem [{\citenamefont {Platini}\ and\ \citenamefont
  {Karevski}(2007)}]{Platini07}%
  \BibitemOpen
  \bibfield  {author} {\bibinfo {author} {\bibfnamefont {T.}~\bibnamefont
  {Platini}}\ and\ \bibinfo {author} {\bibfnamefont {D.}~\bibnamefont
  {Karevski}},\ }\href {http://stacks.iop.org/1751-8121/40/i=8/a=002}
  {\bibfield  {journal} {\bibinfo  {journal} {Journal of Physics A:
  Mathematical and Theoretical}\ }\textbf {\bibinfo {volume} {40}},\ \bibinfo
  {pages} {1711} (\bibinfo {year} {2007})}\BibitemShut {NoStop}%
\bibitem [{\citenamefont {Lancaster}\ and\ \citenamefont
  {Mitra}(2010)}]{Lancaster10}%
  \BibitemOpen
  \bibfield  {author} {\bibinfo {author} {\bibfnamefont {J.}~\bibnamefont
  {Lancaster}}\ and\ \bibinfo {author} {\bibfnamefont {A.}~\bibnamefont
  {Mitra}},\ }\href {\doibase 10.1103/PhysRevE.81.061134} {\bibfield  {journal}
  {\bibinfo  {journal} {Phys. Rev. E}\ }\textbf {\bibinfo {volume} {81}},\
  \bibinfo {pages} {061134} (\bibinfo {year} {2010})}\BibitemShut {NoStop}%
\bibitem [{\citenamefont {Ovchinnikov}(2007)}]{Ovchinnikov07}%
  \BibitemOpen
  \bibfield  {author} {\bibinfo {author} {\bibfnamefont {A.}~\bibnamefont
  {Ovchinnikov}},\ }\href {\doibase
  http://dx.doi.org/10.1016/j.physleta.2007.02.061} {\bibfield  {journal}
  {\bibinfo  {journal} {Physics Letters A}\ }\textbf {\bibinfo {volume}
  {366}},\ \bibinfo {pages} {357 } (\bibinfo {year} {2007})}\BibitemShut
  {NoStop}%
\bibitem [{\citenamefont {Rigol}\ and\ \citenamefont
  {Muramatsu}(2004)}]{Rigol04}%
  \BibitemOpen
  \bibfield  {author} {\bibinfo {author} {\bibfnamefont {M.}~\bibnamefont
  {Rigol}}\ and\ \bibinfo {author} {\bibfnamefont {A.}~\bibnamefont
  {Muramatsu}},\ }\href {\doibase 10.1103/PhysRevLett.93.230404} {\bibfield
  {journal} {\bibinfo  {journal} {Phys. Rev. Lett.}\ }\textbf {\bibinfo
  {volume} {93}},\ \bibinfo {pages} {230404} (\bibinfo {year}
  {2004})}\BibitemShut {NoStop}%
\bibitem [{\citenamefont {Vidmar}\ \emph {et~al.}(2015)\citenamefont {Vidmar},
  \citenamefont {Ronzheimer}, \citenamefont {Schreiber}, \citenamefont {Braun},
  \citenamefont {Hodgman}, \citenamefont {Langer}, \citenamefont
  {Heidrich-Meisner}, \citenamefont {Bloch},\ and\ \citenamefont
  {Schneider}}]{Vidmar15}%
  \BibitemOpen
  \bibfield  {author} {\bibinfo {author} {\bibfnamefont {L.}~\bibnamefont
  {Vidmar}}, \bibinfo {author} {\bibfnamefont {J.~P.}\ \bibnamefont
  {Ronzheimer}}, \bibinfo {author} {\bibfnamefont {M.}~\bibnamefont
  {Schreiber}}, \bibinfo {author} {\bibfnamefont {S.}~\bibnamefont {Braun}},
  \bibinfo {author} {\bibfnamefont {S.~S.}\ \bibnamefont {Hodgman}}, \bibinfo
  {author} {\bibfnamefont {S.}~\bibnamefont {Langer}}, \bibinfo {author}
  {\bibfnamefont {F.}~\bibnamefont {Heidrich-Meisner}}, \bibinfo {author}
  {\bibfnamefont {I.}~\bibnamefont {Bloch}}, \ and\ \bibinfo {author}
  {\bibfnamefont {U.}~\bibnamefont {Schneider}},\ }\href {\doibase
  10.1103/PhysRevLett.115.175301} {\bibfield  {journal} {\bibinfo  {journal}
  {Phys. Rev. Lett.}\ }\textbf {\bibinfo {volume} {115}},\ \bibinfo {pages}
  {175301} (\bibinfo {year} {2015})}\BibitemShut {NoStop}%
\bibitem [{\citenamefont {Langer}\ \emph {et~al.}(2009)\citenamefont {Langer},
  \citenamefont {Heidrich-Meisner}, \citenamefont {Gemmer}, \citenamefont
  {McCulloch},\ and\ \citenamefont {Schollw\"ock}}]{Langer09}%
  \BibitemOpen
  \bibfield  {author} {\bibinfo {author} {\bibfnamefont {S.}~\bibnamefont
  {Langer}}, \bibinfo {author} {\bibfnamefont {F.}~\bibnamefont
  {Heidrich-Meisner}}, \bibinfo {author} {\bibfnamefont {J.}~\bibnamefont
  {Gemmer}}, \bibinfo {author} {\bibfnamefont {I.~P.}\ \bibnamefont
  {McCulloch}}, \ and\ \bibinfo {author} {\bibfnamefont {U.}~\bibnamefont
  {Schollw\"ock}},\ }\href {\doibase 10.1103/PhysRevB.79.214409} {\bibfield
  {journal} {\bibinfo  {journal} {Phys. Rev. B}\ }\textbf {\bibinfo {volume}
  {79}},\ \bibinfo {pages} {214409} (\bibinfo {year} {2009})}\BibitemShut
  {NoStop}%
\bibitem [{\citenamefont {Foster}\ \emph {et~al.}(2011)\citenamefont {Foster},
  \citenamefont {Berkelbach}, \citenamefont {Reichman},\ and\ \citenamefont
  {Yuzbashyan}}]{Foster11}%
  \BibitemOpen
  \bibfield  {author} {\bibinfo {author} {\bibfnamefont {M.~S.}\ \bibnamefont
  {Foster}}, \bibinfo {author} {\bibfnamefont {T.~C.}\ \bibnamefont
  {Berkelbach}}, \bibinfo {author} {\bibfnamefont {D.~R.}\ \bibnamefont
  {Reichman}}, \ and\ \bibinfo {author} {\bibfnamefont {E.~A.}\ \bibnamefont
  {Yuzbashyan}},\ }\href {\doibase 10.1103/PhysRevB.84.085146} {\bibfield
  {journal} {\bibinfo  {journal} {Phys. Rev. B}\ }\textbf {\bibinfo {volume}
  {84}},\ \bibinfo {pages} {085146} (\bibinfo {year} {2011})}\BibitemShut
  {NoStop}%
\bibitem [{\citenamefont {Santos}\ and\ \citenamefont
  {Mitra}(2011)}]{Santos11}%
  \BibitemOpen
  \bibfield  {author} {\bibinfo {author} {\bibfnamefont {L.~F.}\ \bibnamefont
  {Santos}}\ and\ \bibinfo {author} {\bibfnamefont {A.}~\bibnamefont {Mitra}},\
  }\href {\doibase 10.1103/PhysRevE.84.016206} {\bibfield  {journal} {\bibinfo
  {journal} {Phys. Rev. E}\ }\textbf {\bibinfo {volume} {84}},\ \bibinfo
  {pages} {016206} (\bibinfo {year} {2011})}\BibitemShut {NoStop}%
\bibitem [{\citenamefont {Lancaster}\ \emph {et~al.}(2010)\citenamefont
  {Lancaster}, \citenamefont {Gull},\ and\ \citenamefont
  {Mitra}}]{Lancaster10b}%
  \BibitemOpen
  \bibfield  {author} {\bibinfo {author} {\bibfnamefont {J.}~\bibnamefont
  {Lancaster}}, \bibinfo {author} {\bibfnamefont {E.}~\bibnamefont {Gull}}, \
  and\ \bibinfo {author} {\bibfnamefont {A.}~\bibnamefont {Mitra}},\ }\href
  {\doibase 10.1103/PhysRevB.82.235124} {\bibfield  {journal} {\bibinfo
  {journal} {Phys. Rev. B}\ }\textbf {\bibinfo {volume} {82}},\ \bibinfo
  {pages} {235124} (\bibinfo {year} {2010})}\BibitemShut {NoStop}%
\bibitem [{\citenamefont {Vidmar}\ \emph {et~al.}(shed)\citenamefont {Vidmar},
  \citenamefont {Iyer},\ and\ \citenamefont {Rigol}}]{Vidmar16}%
  \BibitemOpen
  \bibfield  {author} {\bibinfo {author} {\bibfnamefont {L.}~\bibnamefont
  {Vidmar}}, \bibinfo {author} {\bibfnamefont {D.}~\bibnamefont {Iyer}}, \ and\
  \bibinfo {author} {\bibfnamefont {M.}~\bibnamefont {Rigol}},\ }\href@noop {}
  {\bibfield  {journal} {\bibinfo  {journal} {arXiv:1512.05373}\ } (\bibinfo
  {year} {unpublished})}\BibitemShut {NoStop}%
\bibitem [{\citenamefont {Bertini}\ \emph {et~al.}(2016)\citenamefont
  {Bertini}, \citenamefont {Collura}, \citenamefont {De~Nardis},\ and\
  \citenamefont {Fagotti}}]{Bertini16}%
  \BibitemOpen
  \bibfield  {author} {\bibinfo {author} {\bibfnamefont {B.}~\bibnamefont
  {Bertini}}, \bibinfo {author} {\bibfnamefont {M.}~\bibnamefont {Collura}},
  \bibinfo {author} {\bibfnamefont {J.}~\bibnamefont {De~Nardis}}, \ and\
  \bibinfo {author} {\bibfnamefont {M.}~\bibnamefont {Fagotti}},\ }\href
  {\doibase 10.1103/PhysRevLett.117.207201} {\bibfield  {journal} {\bibinfo
  {journal} {Phys. Rev. Lett.}\ }\textbf {\bibinfo {volume} {117}},\ \bibinfo
  {pages} {207201} (\bibinfo {year} {2016})}\BibitemShut {NoStop}%
\bibitem [{\citenamefont {Castro-Alvaredo}\ \emph {et~al.}(2016)\citenamefont
  {Castro-Alvaredo}, \citenamefont {Doyon},\ and\ \citenamefont
  {Yoshimura}}]{Doyon16}%
  \BibitemOpen
  \bibfield  {author} {\bibinfo {author} {\bibfnamefont {O.~A.}\ \bibnamefont
  {Castro-Alvaredo}}, \bibinfo {author} {\bibfnamefont {B.}~\bibnamefont
  {Doyon}}, \ and\ \bibinfo {author} {\bibfnamefont {T.}~\bibnamefont
  {Yoshimura}},\ }\href {\doibase 10.1103/PhysRevX.6.041065} {\bibfield
  {journal} {\bibinfo  {journal} {Phys. Rev. X}\ }\textbf {\bibinfo {volume}
  {6}},\ \bibinfo {pages} {041065} (\bibinfo {year} {2016})}\BibitemShut
  {NoStop}%
\bibitem [{\citenamefont {Mazur}(1969)}]{Mazur69}%
  \BibitemOpen
  \bibfield  {author} {\bibinfo {author} {\bibfnamefont {P.}~\bibnamefont
  {Mazur}},\ }\href {\doibase http://dx.doi.org/10.1016/0031-8914(69)90185-2}
  {\bibfield  {journal} {\bibinfo  {journal} {Physica}\ }\textbf {\bibinfo
  {volume} {43}},\ \bibinfo {pages} {533 } (\bibinfo {year}
  {1969})}\BibitemShut {NoStop}%
\bibitem [{\citenamefont {Eisler}\ and\ \citenamefont
  {R\'acz}(2013)}]{Eisler13}%
  \BibitemOpen
  \bibfield  {author} {\bibinfo {author} {\bibfnamefont {V.}~\bibnamefont
  {Eisler}}\ and\ \bibinfo {author} {\bibfnamefont {Z.}~\bibnamefont
  {R\'acz}},\ }\href {\doibase 10.1103/PhysRevLett.110.060602} {\bibfield
  {journal} {\bibinfo  {journal} {Phys. Rev. Lett.}\ }\textbf {\bibinfo
  {volume} {110}},\ \bibinfo {pages} {060602} (\bibinfo {year}
  {2013})}\BibitemShut {NoStop}%
\bibitem [{\citenamefont {Zauner}\ \emph {et~al.}(2015)\citenamefont {Zauner},
  \citenamefont {Ganahl}, \citenamefont {Evertz},\ and\ \citenamefont
  {Nishino}}]{Evertz15}%
  \BibitemOpen
  \bibfield  {author} {\bibinfo {author} {\bibfnamefont {V.}~\bibnamefont
  {Zauner}}, \bibinfo {author} {\bibfnamefont {M.}~\bibnamefont {Ganahl}},
  \bibinfo {author} {\bibfnamefont {H.~G.}\ \bibnamefont {Evertz}}, \ and\
  \bibinfo {author} {\bibfnamefont {T.}~\bibnamefont {Nishino}},\ }\href
  {http://stacks.iop.org/0953-8984/27/i=42/a=425602} {\bibfield  {journal}
  {\bibinfo  {journal} {Journal of Physics: Condensed Matter}\ }\textbf
  {\bibinfo {volume} {27}},\ \bibinfo {pages} {425602} (\bibinfo {year}
  {2015})}\BibitemShut {NoStop}%
\bibitem [{\citenamefont {Sabetta}\ and\ \citenamefont
  {Misguich}(2013)}]{Misguich13}%
  \BibitemOpen
  \bibfield  {author} {\bibinfo {author} {\bibfnamefont {T.}~\bibnamefont
  {Sabetta}}\ and\ \bibinfo {author} {\bibfnamefont {G.}~\bibnamefont
  {Misguich}},\ }\href {\doibase 10.1103/PhysRevB.88.245114} {\bibfield
  {journal} {\bibinfo  {journal} {Phys. Rev. B}\ }\textbf {\bibinfo {volume}
  {88}},\ \bibinfo {pages} {245114} (\bibinfo {year} {2013})}\BibitemShut
  {NoStop}%
\bibitem [{\citenamefont {Cazalilla}(2006)}]{Cazalilla06}%
  \BibitemOpen
  \bibfield  {author} {\bibinfo {author} {\bibfnamefont {M.~A.}\ \bibnamefont
  {Cazalilla}},\ }\href {\doibase 10.1103/PhysRevLett.97.156403} {\bibfield
  {journal} {\bibinfo  {journal} {Phys. Rev. Lett.}\ }\textbf {\bibinfo
  {volume} {97}},\ \bibinfo {pages} {156403} (\bibinfo {year}
  {2006})}\BibitemShut {NoStop}%
\bibitem [{\citenamefont {Iucci}\ and\ \citenamefont
  {Cazalilla}(2009)}]{Iucci09}%
  \BibitemOpen
  \bibfield  {author} {\bibinfo {author} {\bibfnamefont {A.}~\bibnamefont
  {Iucci}}\ and\ \bibinfo {author} {\bibfnamefont {M.~A.}\ \bibnamefont
  {Cazalilla}},\ }\href {\doibase 10.1103/PhysRevA.80.063619} {\bibfield
  {journal} {\bibinfo  {journal} {Phys. Rev. A}\ }\textbf {\bibinfo {volume}
  {80}},\ \bibinfo {pages} {063619} (\bibinfo {year} {2009})}\BibitemShut
  {NoStop}%
\bibitem [{\citenamefont {Mitra}\ and\ \citenamefont
  {Giamarchi}(2011)}]{Mitra11}%
  \BibitemOpen
  \bibfield  {author} {\bibinfo {author} {\bibfnamefont {A.}~\bibnamefont
  {Mitra}}\ and\ \bibinfo {author} {\bibfnamefont {T.}~\bibnamefont
  {Giamarchi}},\ }\href {\doibase 10.1103/PhysRevLett.107.150602} {\bibfield
  {journal} {\bibinfo  {journal} {Phys. Rev. Lett.}\ }\textbf {\bibinfo
  {volume} {107}},\ \bibinfo {pages} {150602} (\bibinfo {year}
  {2011})}\BibitemShut {NoStop}%
\bibitem [{\citenamefont {Karrasch}\ \emph {et~al.}(2012)\citenamefont
  {Karrasch}, \citenamefont {Rentrop}, \citenamefont {Schuricht},\ and\
  \citenamefont {Meden}}]{Karrasch12}%
  \BibitemOpen
  \bibfield  {author} {\bibinfo {author} {\bibfnamefont {C.}~\bibnamefont
  {Karrasch}}, \bibinfo {author} {\bibfnamefont {J.}~\bibnamefont {Rentrop}},
  \bibinfo {author} {\bibfnamefont {D.}~\bibnamefont {Schuricht}}, \ and\
  \bibinfo {author} {\bibfnamefont {V.}~\bibnamefont {Meden}},\ }\href
  {\doibase 10.1103/PhysRevLett.109.126406} {\bibfield  {journal} {\bibinfo
  {journal} {Phys. Rev. Lett.}\ }\textbf {\bibinfo {volume} {109}},\ \bibinfo
  {pages} {126406} (\bibinfo {year} {2012})}\BibitemShut {NoStop}%
\bibitem [{\citenamefont {Kennes}\ and\ \citenamefont
  {Meden}(2013)}]{Kennes13}%
  \BibitemOpen
  \bibfield  {author} {\bibinfo {author} {\bibfnamefont {D.~M.}\ \bibnamefont
  {Kennes}}\ and\ \bibinfo {author} {\bibfnamefont {V.}~\bibnamefont {Meden}},\
  }\href {\doibase 10.1103/PhysRevB.88.165131} {\bibfield  {journal} {\bibinfo
  {journal} {Phys. Rev. B}\ }\textbf {\bibinfo {volume} {88}},\ \bibinfo
  {pages} {165131} (\bibinfo {year} {2013})}\BibitemShut {NoStop}%
\bibitem [{\citenamefont {Collura}\ \emph {et~al.}(2015)\citenamefont
  {Collura}, \citenamefont {Calabrese},\ and\ \citenamefont
  {Essler}}]{Collura15}%
  \BibitemOpen
  \bibfield  {author} {\bibinfo {author} {\bibfnamefont {M.}~\bibnamefont
  {Collura}}, \bibinfo {author} {\bibfnamefont {P.}~\bibnamefont {Calabrese}},
  \ and\ \bibinfo {author} {\bibfnamefont {F.~H.~L.}\ \bibnamefont {Essler}},\
  }\href {\doibase 10.1103/PhysRevB.92.125131} {\bibfield  {journal} {\bibinfo
  {journal} {Phys. Rev. B}\ }\textbf {\bibinfo {volume} {92}},\ \bibinfo
  {pages} {125131} (\bibinfo {year} {2015})}\BibitemShut {NoStop}%
\bibitem [{\citenamefont {Rigol}\ \emph {et~al.}(2007)\citenamefont {Rigol},
  \citenamefont {Dunjko}, \citenamefont {Yurovsky},\ and\ \citenamefont
  {Olshanii}}]{Rigol07}%
  \BibitemOpen
  \bibfield  {author} {\bibinfo {author} {\bibfnamefont {M.}~\bibnamefont
  {Rigol}}, \bibinfo {author} {\bibfnamefont {V.}~\bibnamefont {Dunjko}},
  \bibinfo {author} {\bibfnamefont {V.}~\bibnamefont {Yurovsky}}, \ and\
  \bibinfo {author} {\bibfnamefont {M.}~\bibnamefont {Olshanii}},\ }\href
  {\doibase 10.1103/PhysRevLett.98.050405} {\bibfield  {journal} {\bibinfo
  {journal} {Phys. Rev. Lett.}\ }\textbf {\bibinfo {volume} {98}},\ \bibinfo
  {pages} {050405} (\bibinfo {year} {2007})}\BibitemShut {NoStop}%
\bibitem [{\citenamefont {Schir\'o}\ and\ \citenamefont
  {Mitra}(2014)}]{Schiro14}%
  \BibitemOpen
  \bibfield  {author} {\bibinfo {author} {\bibfnamefont {M.}~\bibnamefont
  {Schir\'o}}\ and\ \bibinfo {author} {\bibfnamefont {A.}~\bibnamefont
  {Mitra}},\ }\href {\doibase 10.1103/PhysRevLett.112.246401} {\bibfield
  {journal} {\bibinfo  {journal} {Phys. Rev. Lett.}\ }\textbf {\bibinfo
  {volume} {112}},\ \bibinfo {pages} {246401} (\bibinfo {year}
  {2014})}\BibitemShut {NoStop}%
\bibitem [{\citenamefont {Schir\'o}\ and\ \citenamefont
  {Mitra}(2015)}]{Schiro15}%
  \BibitemOpen
  \bibfield  {author} {\bibinfo {author} {\bibfnamefont {M.}~\bibnamefont
  {Schir\'o}}\ and\ \bibinfo {author} {\bibfnamefont {A.}~\bibnamefont
  {Mitra}},\ }\href {\doibase 10.1103/PhysRevB.91.235126} {\bibfield  {journal}
  {\bibinfo  {journal} {Phys. Rev. B}\ }\textbf {\bibinfo {volume} {91}},\
  \bibinfo {pages} {235126} (\bibinfo {year} {2015})}\BibitemShut {NoStop}%
\bibitem [{\citenamefont {Caux}\ and\ \citenamefont {Essler}(2013)}]{Caux13}%
  \BibitemOpen
  \bibfield  {author} {\bibinfo {author} {\bibfnamefont {J.-S.}\ \bibnamefont
  {Caux}}\ and\ \bibinfo {author} {\bibfnamefont {F.~H.~L.}\ \bibnamefont
  {Essler}},\ }\href {\doibase 10.1103/PhysRevLett.110.257203} {\bibfield
  {journal} {\bibinfo  {journal} {Phys. Rev. Lett.}\ }\textbf {\bibinfo
  {volume} {110}},\ \bibinfo {pages} {257203} (\bibinfo {year}
  {2013})}\BibitemShut {NoStop}%
\bibitem [{\citenamefont {Ilievski}\ \emph {et~al.}(2015)\citenamefont
  {Ilievski}, \citenamefont {De~Nardis}, \citenamefont {Wouters}, \citenamefont
  {Caux}, \citenamefont {Essler},\ and\ \citenamefont {Prosen}}]{Ilevski15}%
  \BibitemOpen
  \bibfield  {author} {\bibinfo {author} {\bibfnamefont {E.}~\bibnamefont
  {Ilievski}}, \bibinfo {author} {\bibfnamefont {J.}~\bibnamefont {De~Nardis}},
  \bibinfo {author} {\bibfnamefont {B.}~\bibnamefont {Wouters}}, \bibinfo
  {author} {\bibfnamefont {J.-S.}\ \bibnamefont {Caux}}, \bibinfo {author}
  {\bibfnamefont {F.~H.~L.}\ \bibnamefont {Essler}}, \ and\ \bibinfo {author}
  {\bibfnamefont {T.}~\bibnamefont {Prosen}},\ }\href {\doibase
  10.1103/PhysRevLett.115.157201} {\bibfield  {journal} {\bibinfo  {journal}
  {Phys. Rev. Lett.}\ }\textbf {\bibinfo {volume} {115}},\ \bibinfo {pages}
  {157201} (\bibinfo {year} {2015})}\BibitemShut {NoStop}%
\bibitem [{\citenamefont {Caux}(2016)}]{Caux16}%
  \BibitemOpen
  \bibfield  {author} {\bibinfo {author} {\bibfnamefont {J.-S.}\ \bibnamefont
  {Caux}},\ }\href {http://stacks.iop.org/1742-5468/2016/i=6/a=064006}
  {\bibfield  {journal} {\bibinfo  {journal} {Journal of Statistical Mechanics:
  Theory and Experiment}\ ,\ \bibinfo {pages} {064006}} (\bibinfo {year}
  {2016})}\BibitemShut {NoStop}%
\bibitem [{\citenamefont {Tavora}\ \emph {et~al.}(2014)\citenamefont {Tavora},
  \citenamefont {Rosch},\ and\ \citenamefont {Mitra}}]{Tavora14}%
  \BibitemOpen
  \bibfield  {author} {\bibinfo {author} {\bibfnamefont {M.}~\bibnamefont
  {Tavora}}, \bibinfo {author} {\bibfnamefont {A.}~\bibnamefont {Rosch}}, \
  and\ \bibinfo {author} {\bibfnamefont {A.}~\bibnamefont {Mitra}},\ }\href
  {\doibase 10.1103/PhysRevLett.113.010601} {\bibfield  {journal} {\bibinfo
  {journal} {Phys. Rev. Lett.}\ }\textbf {\bibinfo {volume} {113}},\ \bibinfo
  {pages} {010601} (\bibinfo {year} {2014})}\BibitemShut {NoStop}%
\bibitem [{\citenamefont {Singh}\ \emph {et~al.}(1989)\citenamefont {Singh},
  \citenamefont {Fisher},\ and\ \citenamefont {Shankar}}]{Singh89}%
  \BibitemOpen
  \bibfield  {author} {\bibinfo {author} {\bibfnamefont {R.~R.~P.}\
  \bibnamefont {Singh}}, \bibinfo {author} {\bibfnamefont {M.~E.}\ \bibnamefont
  {Fisher}}, \ and\ \bibinfo {author} {\bibfnamefont {R.}~\bibnamefont
  {Shankar}},\ }\href {\doibase 10.1103/PhysRevB.39.2562} {\bibfield  {journal}
  {\bibinfo  {journal} {Phys. Rev. B}\ }\textbf {\bibinfo {volume} {39}},\
  \bibinfo {pages} {2562} (\bibinfo {year} {1989})}\BibitemShut {NoStop}%
\bibitem [{\citenamefont {Affleck}(1998)}]{Affleck98}%
  \BibitemOpen
  \bibfield  {author} {\bibinfo {author} {\bibfnamefont {I.}~\bibnamefont
  {Affleck}},\ }\href@noop {} {\bibfield  {journal} {\bibinfo  {journal} {J.
  Phys. A: Math. Gen.}\ }\textbf {\bibinfo {volume} {31}},\ \bibinfo {pages}
  {4573} (\bibinfo {year} {1998})}\BibitemShut {NoStop}%
\bibitem [{\citenamefont {Mitra}(2013)}]{Mitra13a}%
  \BibitemOpen
  \bibfield  {author} {\bibinfo {author} {\bibfnamefont {A.}~\bibnamefont
  {Mitra}},\ }\href {\doibase 10.1103/PhysRevB.87.205109} {\bibfield  {journal}
  {\bibinfo  {journal} {Phys. Rev. B}\ }\textbf {\bibinfo {volume} {87}},\
  \bibinfo {pages} {205109} (\bibinfo {year} {2013})}\BibitemShut {NoStop}%
\bibitem [{\citenamefont {Mitra}\ and\ \citenamefont
  {Giamarchi}(2012)}]{Mitra12a}%
  \BibitemOpen
  \bibfield  {author} {\bibinfo {author} {\bibfnamefont {A.}~\bibnamefont
  {Mitra}}\ and\ \bibinfo {author} {\bibfnamefont {T.}~\bibnamefont
  {Giamarchi}},\ }\href {\doibase 10.1103/PhysRevB.85.075117} {\bibfield
  {journal} {\bibinfo  {journal} {Phys. Rev. B}\ }\textbf {\bibinfo {volume}
  {85}},\ \bibinfo {pages} {075117} (\bibinfo {year} {2012})}\BibitemShut
  {NoStop}%
\bibitem [{\citenamefont {Mitra}(2012)}]{Mitra12b}%
  \BibitemOpen
  \bibfield  {author} {\bibinfo {author} {\bibfnamefont {A.}~\bibnamefont
  {Mitra}},\ }\href {\doibase 10.1103/PhysRevLett.109.260601} {\bibfield
  {journal} {\bibinfo  {journal} {Phys. Rev. Lett.}\ }\textbf {\bibinfo
  {volume} {109}},\ \bibinfo {pages} {260601} (\bibinfo {year}
  {2012})}\BibitemShut {NoStop}%
\bibitem [{\citenamefont {Kamenev}(2011)}]{Kamenevbook2011}%
  \BibitemOpen
  \bibfield  {author} {\bibinfo {author} {\bibfnamefont {A.}~\bibnamefont
  {Kamenev}},\ }\href@noop {} {\emph {\bibinfo {title} {Field Theory of
  Non-Equilibrium Systems}}}\ (\bibinfo  {publisher} {Cambridge University
  Press},\ \bibinfo {year} {2011})\BibitemShut {NoStop}%
\bibitem [{\citenamefont {Berges}(2004)}]{BergesRev}%
  \BibitemOpen
  \bibfield  {author} {\bibinfo {author} {\bibfnamefont {J.}~\bibnamefont
  {Berges}},\ }\href@noop {} {\bibfield  {journal} {\bibinfo  {journal} {AIP
  Conf. Proc. 739}\ ,\ \bibinfo {pages} {3}} (\bibinfo {year}
  {2004})}\BibitemShut {NoStop}%
\bibitem [{\citenamefont {Cornwall}\ \emph {et~al.}(1974)\citenamefont
  {Cornwall}, \citenamefont {Jackiw},\ and\ \citenamefont
  {Tomboulis}}]{CornwallJackiwTomboulis}%
  \BibitemOpen
  \bibfield  {author} {\bibinfo {author} {\bibfnamefont {J.~M.}\ \bibnamefont
  {Cornwall}}, \bibinfo {author} {\bibfnamefont {R.}~\bibnamefont {Jackiw}}, \
  and\ \bibinfo {author} {\bibfnamefont {E.}~\bibnamefont {Tomboulis}},\
  }\href@noop {} {\bibfield  {journal} {\bibinfo  {journal} {Phys. Rev. D}\
  }\textbf {\bibinfo {volume} {10}},\ \bibinfo {pages} {2428} (\bibinfo {year}
  {1974})}\BibitemShut {NoStop}%
\bibitem [{\citenamefont {Nishiyama}(2010)}]{Nishiyama}%
  \BibitemOpen
  \bibfield  {author} {\bibinfo {author} {\bibfnamefont {A.}~\bibnamefont
  {Nishiyama}},\ }\href@noop {} {\bibfield  {journal} {\bibinfo  {journal}
  {Nuclear Physics A}\ }\textbf {\bibinfo {volume} {832}},\ \bibinfo {pages}
  {289} (\bibinfo {year} {2010})}\BibitemShut {NoStop}%
\bibitem [{\citenamefont {Arrizabalaga}\ \emph {et~al.}(2005)\citenamefont
  {Arrizabalaga}, \citenamefont {Smit},\ and\ \citenamefont
  {Tranberg}}]{AST61}%
  \BibitemOpen
  \bibfield  {author} {\bibinfo {author} {\bibfnamefont {A.}~\bibnamefont
  {Arrizabalaga}}, \bibinfo {author} {\bibfnamefont {J.}~\bibnamefont {Smit}},
  \ and\ \bibinfo {author} {\bibfnamefont {A.}~\bibnamefont {Tranberg}},\
  }\href@noop {} {\bibfield  {journal} {\bibinfo  {journal} {Phys. Rev. D}\
  }\textbf {\bibinfo {volume} {72}},\ \bibinfo {pages} {025014} (\bibinfo
  {year} {2005})}\BibitemShut {NoStop}%
\bibitem [{\citenamefont {Ivanov}\ \emph {et~al.}(1999)\citenamefont {Ivanov},
  \citenamefont {Knoll},\ and\ \citenamefont {Voskresensky}}]{Ivanov98}%
  \BibitemOpen
  \bibfield  {author} {\bibinfo {author} {\bibfnamefont {Y.~B.}\ \bibnamefont
  {Ivanov}}, \bibinfo {author} {\bibfnamefont {J.}~\bibnamefont {Knoll}}, \
  and\ \bibinfo {author} {\bibfnamefont {D.~N.}\ \bibnamefont {Voskresensky}},\
  }\href@noop {} {\bibfield  {journal} {\bibinfo  {journal} {Nuclear Physics
  A}\ }\textbf {\bibinfo {volume} {657}},\ \bibinfo {pages} {413} (\bibinfo
  {year} {1999})}\BibitemShut {NoStop}%
\bibitem [{\citenamefont {Tavora}\ and\ \citenamefont
  {Mitra}(2013)}]{Tavora13}%
  \BibitemOpen
  \bibfield  {author} {\bibinfo {author} {\bibfnamefont {M.}~\bibnamefont
  {Tavora}}\ and\ \bibinfo {author} {\bibfnamefont {A.}~\bibnamefont {Mitra}},\
  }\href {\doibase 10.1103/PhysRevB.88.115144} {\bibfield  {journal} {\bibinfo
  {journal} {Phys. Rev. B}\ }\textbf {\bibinfo {volume} {88}},\ \bibinfo
  {pages} {115144} (\bibinfo {year} {2013})}\BibitemShut {NoStop}%
\bibitem [{\citenamefont {Grams}\ \emph {et~al.}(2014)\citenamefont {Grams},
  \citenamefont {Valldor}, \citenamefont {Garst},\ and\ \citenamefont
  {Hemberger}}]{Grams14}%
  \BibitemOpen
  \bibfield  {author} {\bibinfo {author} {\bibfnamefont {C.~P.}\ \bibnamefont
  {Grams}}, \bibinfo {author} {\bibfnamefont {M.}~\bibnamefont {Valldor}},
  \bibinfo {author} {\bibfnamefont {M.}~\bibnamefont {Garst}}, \ and\ \bibinfo
  {author} {\bibfnamefont {J.}~\bibnamefont {Hemberger}},\ }\href@noop {}
  {\bibfield  {journal} {\bibinfo  {journal} {Nature Comm.}\ }\textbf {\bibinfo
  {volume} {5}},\ \bibinfo {pages} {4853} (\bibinfo {year} {2014})}\BibitemShut
  {NoStop}%
\bibitem [{\citenamefont {Lux}\ \emph {et~al.}(2014)\citenamefont {Lux},
  \citenamefont {M\"uller}, \citenamefont {Mitra},\ and\ \citenamefont
  {Rosch}}]{Lux13}%
  \BibitemOpen
  \bibfield  {author} {\bibinfo {author} {\bibfnamefont {J.}~\bibnamefont
  {Lux}}, \bibinfo {author} {\bibfnamefont {J.}~\bibnamefont {M\"uller}},
  \bibinfo {author} {\bibfnamefont {A.}~\bibnamefont {Mitra}}, \ and\ \bibinfo
  {author} {\bibfnamefont {A.}~\bibnamefont {Rosch}},\ }\href {\doibase
  10.1103/PhysRevA.89.053608} {\bibfield  {journal} {\bibinfo  {journal} {Phys.
  Rev. A}\ }\textbf {\bibinfo {volume} {89}},\ \bibinfo {pages} {053608}
  (\bibinfo {year} {2014})}\BibitemShut {NoStop}%
\bibitem [{\citenamefont {Kardar}\ \emph {et~al.}(1986)\citenamefont {Kardar},
  \citenamefont {Parisi},\ and\ \citenamefont {Zhang}}]{KPZ86}%
  \BibitemOpen
  \bibfield  {author} {\bibinfo {author} {\bibfnamefont {M.}~\bibnamefont
  {Kardar}}, \bibinfo {author} {\bibfnamefont {G.}~\bibnamefont {Parisi}}, \
  and\ \bibinfo {author} {\bibfnamefont {Y.-C.}\ \bibnamefont {Zhang}},\ }\href
  {\doibase 10.1103/PhysRevLett.56.889} {\bibfield  {journal} {\bibinfo
  {journal} {Phys. Rev. Lett.}\ }\textbf {\bibinfo {volume} {56}},\ \bibinfo
  {pages} {889} (\bibinfo {year} {1986})}\BibitemShut {NoStop}%
\bibitem [{\citenamefont {van Beijeren}(2012)}]{Henk12}%
  \BibitemOpen
  \bibfield  {author} {\bibinfo {author} {\bibfnamefont {H.}~\bibnamefont {van
  Beijeren}},\ }\href {\doibase 10.1103/PhysRevLett.108.180601} {\bibfield
  {journal} {\bibinfo  {journal} {Phys. Rev. Lett.}\ }\textbf {\bibinfo
  {volume} {108}},\ \bibinfo {pages} {180601} (\bibinfo {year}
  {2012})}\BibitemShut {NoStop}%
\bibitem [{\citenamefont {Schir\'o}\ and\ \citenamefont
  {Fabrizio}(2010)}]{Schiro10}%
  \BibitemOpen
  \bibfield  {author} {\bibinfo {author} {\bibfnamefont {M.}~\bibnamefont
  {Schir\'o}}\ and\ \bibinfo {author} {\bibfnamefont {M.}~\bibnamefont
  {Fabrizio}},\ }\href {\doibase 10.1103/PhysRevLett.105.076401} {\bibfield
  {journal} {\bibinfo  {journal} {Phys. Rev. Lett.}\ }\textbf {\bibinfo
  {volume} {105}},\ \bibinfo {pages} {076401} (\bibinfo {year}
  {2010})}\BibitemShut {NoStop}%
\bibitem [{\citenamefont {Gambassi}\ and\ \citenamefont
  {Calabrese}(2011)}]{Gambassi2011}%
  \BibitemOpen
  \bibfield  {author} {\bibinfo {author} {\bibfnamefont {A.}~\bibnamefont
  {Gambassi}}\ and\ \bibinfo {author} {\bibfnamefont {P.}~\bibnamefont
  {Calabrese}},\ }\href {http://stacks.iop.org/0295-5075/95/i=6/a=66007}
  {\bibfield  {journal} {\bibinfo  {journal} {Europhys. Lett.}\ }\textbf
  {\bibinfo {volume} {95}},\ \bibinfo {pages} {66007} (\bibinfo {year}
  {2011})}\BibitemShut {NoStop}%
\bibitem [{\citenamefont {Sciolla}\ and\ \citenamefont
  {Biroli}(2013)}]{Sciolla2013}%
  \BibitemOpen
  \bibfield  {author} {\bibinfo {author} {\bibfnamefont {B.}~\bibnamefont
  {Sciolla}}\ and\ \bibinfo {author} {\bibfnamefont {G.}~\bibnamefont
  {Biroli}},\ }\href {\doibase 10.1103/PhysRevB.88.201110} {\bibfield
  {journal} {\bibinfo  {journal} {Phys. Rev. B}\ }\textbf {\bibinfo {volume}
  {88}},\ \bibinfo {pages} {201110} (\bibinfo {year} {2013})}\BibitemShut
  {NoStop}%
\bibitem [{\citenamefont {Chandran}\ \emph {et~al.}(2013)\citenamefont
  {Chandran}, \citenamefont {Nanduri}, \citenamefont {Gubser},\ and\
  \citenamefont {Sondhi}}]{Sondhi2013}%
  \BibitemOpen
  \bibfield  {author} {\bibinfo {author} {\bibfnamefont {A.}~\bibnamefont
  {Chandran}}, \bibinfo {author} {\bibfnamefont {A.}~\bibnamefont {Nanduri}},
  \bibinfo {author} {\bibfnamefont {S.~S.}\ \bibnamefont {Gubser}}, \ and\
  \bibinfo {author} {\bibfnamefont {S.~L.}\ \bibnamefont {Sondhi}},\ }\href
  {\doibase 10.1103/PhysRevB.88.024306} {\bibfield  {journal} {\bibinfo
  {journal} {Phys. Rev. B}\ }\textbf {\bibinfo {volume} {88}},\ \bibinfo
  {pages} {024306} (\bibinfo {year} {2013})}\BibitemShut {NoStop}%
\bibitem [{\citenamefont {Smacchia}\ \emph {et~al.}(2015)\citenamefont
  {Smacchia}, \citenamefont {Knap}, \citenamefont {Demler},\ and\ \citenamefont
  {Silva}}]{Smacchia2014}%
  \BibitemOpen
  \bibfield  {author} {\bibinfo {author} {\bibfnamefont {P.}~\bibnamefont
  {Smacchia}}, \bibinfo {author} {\bibfnamefont {M.}~\bibnamefont {Knap}},
  \bibinfo {author} {\bibfnamefont {E.}~\bibnamefont {Demler}}, \ and\ \bibinfo
  {author} {\bibfnamefont {A.}~\bibnamefont {Silva}},\ }\href {\doibase
  10.1103/PhysRevB.91.205136} {\bibfield  {journal} {\bibinfo  {journal} {Phys.
  Rev. B}\ }\textbf {\bibinfo {volume} {91}},\ \bibinfo {pages} {205136}
  (\bibinfo {year} {2015})}\BibitemShut {NoStop}%
\bibitem [{\citenamefont {Chiocchetta}\ \emph {et~al.}(2015)\citenamefont
  {Chiocchetta}, \citenamefont {Tavora}, \citenamefont {Gambassi},\ and\
  \citenamefont {Mitra}}]{Chiocchetta2015}%
  \BibitemOpen
  \bibfield  {author} {\bibinfo {author} {\bibfnamefont {A.}~\bibnamefont
  {Chiocchetta}}, \bibinfo {author} {\bibfnamefont {M.}~\bibnamefont {Tavora}},
  \bibinfo {author} {\bibfnamefont {A.}~\bibnamefont {Gambassi}}, \ and\
  \bibinfo {author} {\bibfnamefont {A.}~\bibnamefont {Mitra}},\ }\href
  {\doibase 10.1103/PhysRevB.91.220302} {\bibfield  {journal} {\bibinfo
  {journal} {Phys. Rev. B}\ }\textbf {\bibinfo {volume} {91}},\ \bibinfo
  {pages} {220302} (\bibinfo {year} {2015})}\BibitemShut {NoStop}%
\bibitem [{\citenamefont {Maraga}\ \emph {et~al.}(2015)\citenamefont {Maraga},
  \citenamefont {Chiocchetta}, \citenamefont {Mitra},\ and\ \citenamefont
  {Gambassi}}]{Maraga2015}%
  \BibitemOpen
  \bibfield  {author} {\bibinfo {author} {\bibfnamefont {A.}~\bibnamefont
  {Maraga}}, \bibinfo {author} {\bibfnamefont {A.}~\bibnamefont {Chiocchetta}},
  \bibinfo {author} {\bibfnamefont {A.}~\bibnamefont {Mitra}}, \ and\ \bibinfo
  {author} {\bibfnamefont {A.}~\bibnamefont {Gambassi}},\ }\href {\doibase
  10.1103/PhysRevE.92.042151} {\bibfield  {journal} {\bibinfo  {journal} {Phys.
  Rev. E}\ }\textbf {\bibinfo {volume} {92}},\ \bibinfo {pages} {042151}
  (\bibinfo {year} {2015})}\BibitemShut {NoStop}%
\bibitem [{\citenamefont {Chiocchetta}\ \emph {et~al.}(2016)\citenamefont
  {Chiocchetta}, \citenamefont {Tavora}, \citenamefont {Gambassi},\ and\
  \citenamefont {Mitra}}]{MitGam16}%
  \BibitemOpen
  \bibfield  {author} {\bibinfo {author} {\bibfnamefont {A.}~\bibnamefont
  {Chiocchetta}}, \bibinfo {author} {\bibfnamefont {M.}~\bibnamefont {Tavora}},
  \bibinfo {author} {\bibfnamefont {A.}~\bibnamefont {Gambassi}}, \ and\
  \bibinfo {author} {\bibfnamefont {A.}~\bibnamefont {Mitra}},\ }\href
  {\doibase 10.1103/PhysRevB.94.134311} {\bibfield  {journal} {\bibinfo
  {journal} {Phys. Rev. B}\ }\textbf {\bibinfo {volume} {94}},\ \bibinfo
  {pages} {134311} (\bibinfo {year} {2016})}\BibitemShut {NoStop}%
\bibitem [{\citenamefont {Chiocchetta}\ \emph {et~al.}(2017)\citenamefont
  {Chiocchetta}, \citenamefont {Gambassi}, \citenamefont {Diehl},\ and\
  \citenamefont {Marino}}]{Gambassi17}%
  \BibitemOpen
  \bibfield  {author} {\bibinfo {author} {\bibfnamefont {A.}~\bibnamefont
  {Chiocchetta}}, \bibinfo {author} {\bibfnamefont {A.}~\bibnamefont
  {Gambassi}}, \bibinfo {author} {\bibfnamefont {S.}~\bibnamefont {Diehl}}, \
  and\ \bibinfo {author} {\bibfnamefont {J.}~\bibnamefont {Marino}},\ }\href
  {\doibase 10.1103/PhysRevLett.118.135701} {\bibfield  {journal} {\bibinfo
  {journal} {Phys. Rev. Lett.}\ }\textbf {\bibinfo {volume} {118}},\ \bibinfo
  {pages} {135701} (\bibinfo {year} {2017})}\BibitemShut {NoStop}%
\bibitem [{\citenamefont {Janssen}\ \emph {et~al.}(1989)\citenamefont
  {Janssen}, \citenamefont {Schaub},\ and\ \citenamefont
  {Schmittmann}}]{Janssen1988}%
  \BibitemOpen
  \bibfield  {author} {\bibinfo {author} {\bibfnamefont {H.}~\bibnamefont
  {Janssen}}, \bibinfo {author} {\bibfnamefont {B.}~\bibnamefont {Schaub}}, \
  and\ \bibinfo {author} {\bibfnamefont {B.}~\bibnamefont {Schmittmann}},\
  }\href {\doibase 10.1007/BF01319383} {\bibfield  {journal} {\bibinfo
  {journal} {Z. Phys. B}\ }\textbf {\bibinfo {volume} {73}},\ \bibinfo {pages}
  {539} (\bibinfo {year} {1989})}\BibitemShut {NoStop}%
\bibitem [{\citenamefont {Huse}(1989)}]{Huse89}%
  \BibitemOpen
  \bibfield  {author} {\bibinfo {author} {\bibfnamefont {D.~A.}\ \bibnamefont
  {Huse}},\ }\href {\doibase 10.1103/PhysRevB.40.304} {\bibfield  {journal}
  {\bibinfo  {journal} {Phys. Rev. B}\ }\textbf {\bibinfo {volume} {40}},\
  \bibinfo {pages} {304} (\bibinfo {year} {1989})}\BibitemShut {NoStop}%
\bibitem [{\citenamefont {Calabrese}\ and\ \citenamefont
  {Gambassi}(2005)}]{Gambassi05}%
  \BibitemOpen
  \bibfield  {author} {\bibinfo {author} {\bibfnamefont {P.}~\bibnamefont
  {Calabrese}}\ and\ \bibinfo {author} {\bibfnamefont {A.}~\bibnamefont
  {Gambassi}},\ }\href {http://stacks.iop.org/0305-4470/38/i=18/a=R01}
  {\bibfield  {journal} {\bibinfo  {journal} {Journal of Physics A:
  Mathematical and General}\ }\textbf {\bibinfo {volume} {38}},\ \bibinfo
  {pages} {R133} (\bibinfo {year} {2005})}\BibitemShut {NoStop}%
\bibitem [{\citenamefont {Diehl}(1986)}]{Diehlbook}%
  \BibitemOpen
  \bibfield  {author} {\bibinfo {author} {\bibfnamefont {H.~W.}\ \bibnamefont
  {Diehl}},\ }\href@noop {} {\emph {\bibinfo {title} {Phase Transitions and
  Critical Phenomena}}}\ (\bibinfo  {publisher} {Academic Press, London},\
  \bibinfo {year} {1986})\ \bibinfo {note} {, edited by C. Domb and J. L.
  Lebowitz}\BibitemShut {NoStop}%
\bibitem [{\citenamefont {Lemonik}\ and\ \citenamefont
  {Mitra}(2016)}]{Lemonik16}%
  \BibitemOpen
  \bibfield  {author} {\bibinfo {author} {\bibfnamefont {Y.}~\bibnamefont
  {Lemonik}}\ and\ \bibinfo {author} {\bibfnamefont {A.}~\bibnamefont
  {Mitra}},\ }\href {\doibase 10.1103/PhysRevB.94.024306} {\bibfield  {journal}
  {\bibinfo  {journal} {Phys. Rev. B}\ }\textbf {\bibinfo {volume} {94}},\
  \bibinfo {pages} {024306} (\bibinfo {year} {2016})}\BibitemShut {NoStop}%
\bibitem [{\citenamefont {Peschel}\ and\ \citenamefont
  {Eisler}(2009)}]{Eisler2009}%
  \BibitemOpen
  \bibfield  {author} {\bibinfo {author} {\bibfnamefont {I.}~\bibnamefont
  {Peschel}}\ and\ \bibinfo {author} {\bibfnamefont {V.}~\bibnamefont
  {Eisler}},\ }\href {http://stacks.iop.org/1751-8121/42/i=50/a=504003}
  {\bibfield  {journal} {\bibinfo  {journal} {Journal of Physics A:
  Mathematical and Theoretical}\ }\textbf {\bibinfo {volume} {42}},\ \bibinfo
  {pages} {504003} (\bibinfo {year} {2009})}\BibitemShut {NoStop}%
\bibitem [{\citenamefont {Mitrano}\ \emph {et~al.}(2016)\citenamefont
  {Mitrano}, \citenamefont {Cantaluppi}, \citenamefont {Nicoletti},
  \citenamefont {Kaiser}, \citenamefont {Perucchi}, \citenamefont {Lupi},
  \citenamefont {Pietro}, \citenamefont {Pontiroli}, \citenamefont {Ricc\'{o}},
  \citenamefont {Clark}, \citenamefont {Jaksch},\ and\ \citenamefont
  {Cavalleri}}]{Mitrano15}%
  \BibitemOpen
  \bibfield  {author} {\bibinfo {author} {\bibfnamefont {M.}~\bibnamefont
  {Mitrano}}, \bibinfo {author} {\bibfnamefont {A.}~\bibnamefont {Cantaluppi}},
  \bibinfo {author} {\bibfnamefont {D.}~\bibnamefont {Nicoletti}}, \bibinfo
  {author} {\bibfnamefont {S.}~\bibnamefont {Kaiser}}, \bibinfo {author}
  {\bibfnamefont {A.}~\bibnamefont {Perucchi}}, \bibinfo {author}
  {\bibfnamefont {S.}~\bibnamefont {Lupi}}, \bibinfo {author} {\bibfnamefont
  {P.~D.}\ \bibnamefont {Pietro}}, \bibinfo {author} {\bibfnamefont
  {D.}~\bibnamefont {Pontiroli}}, \bibinfo {author} {\bibfnamefont
  {M.}~\bibnamefont {Ricc\'{o}}}, \bibinfo {author} {\bibfnamefont {S.~R.}\
  \bibnamefont {Clark}}, \bibinfo {author} {\bibfnamefont {D.}~\bibnamefont
  {Jaksch}}, \ and\ \bibinfo {author} {\bibfnamefont {A.}~\bibnamefont
  {Cavalleri}},\ }\href@noop {} {\bibfield  {journal} {\bibinfo  {journal}
  {Nature}\ }\textbf {\bibinfo {volume} {530}},\ \bibinfo {pages} {461}
  (\bibinfo {year} {2016})}\BibitemShut {NoStop}%
\bibitem [{\citenamefont {Dehghani}\ and\ \citenamefont
  {Mitra}(shed)}]{Dehghani17}%
  \BibitemOpen
  \bibfield  {author} {\bibinfo {author} {\bibfnamefont {H.}~\bibnamefont
  {Dehghani}}\ and\ \bibinfo {author} {\bibfnamefont {A.}~\bibnamefont
  {Mitra}},\ }\href@noop {} {\bibfield  {journal} {\bibinfo  {journal}
  {arXiv:1703.01621}\ } (\bibinfo {year} {unpublished})}\BibitemShut {NoStop}%
\end{thebibliography}

\end{document}